\documentstyle[aps,preprint,tighten,epsf]{revtex}
\begin{document}
\preprint{\vbox{\hfill SHEP 96/28\\
                \null\hfill HLRZ 03/97\\
                \null\hfill WUB 97--11\\ }}

\title
{Complete ${\mathcal O}(v^2)$ corrections to the
static interquark
potential from SU(3) gauge theory}
\author{Gunnar S.\ Bali\thanks{Electronic address: bali@hep.ph.soton.ac.uk}}
\address{Department of Physics, The University of
Southampton, Highfield, Southampton SO17 1BJ, England}
\author{Klaus Schilling\thanks{Electronic address:
schillin@theorie.physik.uni-wuppertal.de} and Armin
Wachter\thanks{Electronic address: wachter@hlrserv.hlrz.kfa-juelich.de}}
\address{HLRZ, Forschungszentrum J\"ulich, D-52425 J\"ulich and
DESY, D-22603 Hamburg, Germany\\ and
Fachbereich Physik, Bergische Universit\"at, Gesamthochschule
Wuppertal, Gau\ss{}stra\ss{}e 20, 42097 Wuppertal, Germany}
%\date{\today}
\maketitle

\begin{abstract}
For the first time, we determine the complete spin- and momentum-dependent
order $v^2$ corrections to the static interquark potential from
simulations of QCD in the valence quark approximation
at inverse lattice spacings of 2--3~GeV.
A new flavor dependent correction to the central potential is
found. We report a $r^{-2}$
contribution to the long range spin-orbit
potential $V_1'$. The other spin-dependent potentials turn
out to be short ranged and can be
well understood by means of perturbation theory. The
momentum-dependent potentials qualitatively agree with minimal area
law expectations. In view of spectrum calculations, we discuss the
matching of the effective nonrelativistic theory
to QCD as well as renormalization of lattice results.
In a first survey of the resulting bottomonia and charmonia
spectra we reproduce the experimental levels within
average errors of $12.5$~MeV and
$22$~MeV, respectively.
\end{abstract}
 
\narrowtext
\section{Introduction}
Quarkonia spectroscopy provides a wealth of information and thus
constitutes an important observational window to the phenomenology of
confining quark interactions.
It has been known for a long time that
purely phenomenological or QCD inspired potential models offer  a
suitable heuristic framework to understand the empirical  charmonium
($J/\psi$) and bottomonium ($\Upsilon$)
spectra~\cite{cornell,quigg,lucha,fulcher}.

On a more fundamental level, one would prefer to start out from the
basic QCD Lagrangian to solve the heavy quarkonia bound state problem.
NRQCD~\cite{NRQCD} offers a systematic way to solve this
problem by direct extraction of bound state masses from {\it
effective nonrelativistic} lattice Lagrangians, which approximate the
QCD Lagrangian to a given order in the quark velocity, $v$.
Considerable success has been achieved recently in determining
quarkonia spectra within this approximation to QCD~\cite{nrqcd}.

Here, we follow a complementary strategy: instead of separately computing
the spectral properties of individual mesonic
states, we integrate out the gauge background and directly
determine the underlying quantum mechanical two-particle
Hamiltonian. Once QCD binding problems are recast into
this form, spectra,
wave functions and decay constants for arbitrary (sufficiently large)
quark masses and quantum numbers
can easily be obtained.
Results can either be
confronted with experiment or compared to
predictions from lattice NRQCD.

In the limit of infinite quark mass,
the Born-Oppenheimer approximation is applicable
and, after integrating out the gauge degrees of freedom,
QCD binding problems become nonrelativistic. The static interaction
potential can be computed directly from the QCD Lagrangian on the lattice.
Within the present study, we find
the average velocity between the sources
to be $\langle v^2\rangle\approx 0.27$
and $\langle v^2\rangle\approx 0.08$ for the charmonium and
bottomonium ground-states, respectively. This leads
us to expect that the phenomenological
potentials within those models, which have been
optimized to reproduce empirical spectra,
should deviate by substantial
${\mathcal O}(v^2)$ corrections
{}from the static potential as predicted by QCD;
at realistic quark masses such corrections,
that are also required to obtain hyperfine splittings,
cannot be neglected. Therefore, we have to take
corrections to the static limit into account.

The Hamiltonian that we derive
is equivalent to the QCD
Lagrangian up to ${\mathcal O}(v^2)$. It includes
the spin-dependent (SD) terms derived by Eichten, Feinberg and
Gromes~\cite{eich-fein,gromes},
the momentum-dependent (MD)
corrections derived by Barchielli, Brambilla, Montaldi and Prosperi
(BBMP)~\cite{BBMP} and one-loop radiative corrections
from matching the effective theory to the full theory at a scale $\mu$
that, in general, differs from the heavy quark mass, $m$~\cite{Chen}.
It can be parametrized in terms
of seven independent scalar functions of the quark separation
(the potentials). These will be computed nonperturbatively on the lattice.

The static potential has been determined with high accuracy
in the valence quark (quenched) approximation to
QCD~\cite{pot,pot2,michael3} and, more recently, in full
QCD with two dynamical flavors of light Wilson sea quarks~\cite{SESAM}.
First attempts to compute relativistic corrections have
been made in the mid
eighties for SU(2) and SU(3) gauge
theory~\cite{michael1,rebbi,forcrand,huntley} and have been extended
to QCD with sea quarks in Refs.~\cite{koike,laermann}.

In view of
the general interest in the Hamiltonian formulation of the meson binding
problem, renewed effort should be made to unravel the
structure of the SD potentials and other ${\mathcal
O}(v^2)$ corrections.
Recently, we presented
improved techniques for computation of SD corrections
and tested them successfully
on SU(2) gauge theory~\cite{wachter}.
Here, we shall apply these methods to the
physically relevant SU(3) gauge theory. We will extend the SU(2)
investigation by inclusion of MD potentials and relativistic
corrections to the central potential, and subsequently determine
quarkonia levels.

We wish to emphasize that the method presented does
not rely on any other approximations than truncating
the QCD Lagrangian at second order in the quark velocity (apart from
the valence quark approximation). However, the Schr\"odinger-Pauli
approach to heavy quark binding problems
suffers from the same difficulties as NRQCD, 
namely (a) the error involved in truncating the expansion
at a finite order in $v$, (b) uncertainties in the matching
of the effective Hamiltonian to QCD and (c) renormalization of
lattice results.
While we manage to solve the
latter problem in a satisfactory way, we have to rely on one-loop
perturbation theory for the matching of the nonrelativistic
Hamiltonian to QCD.
Systematic errors from the ${\mathcal O}(v^2)$ approximation as well as
from the uncertainty in the matching constants (that can be
reduced order by order in perturbation theory) are estimated.

Since NRQCD to order $v^2$ (or $v^4$, depending on the labeling
conventions used) is based on the same Lagrangian, it is
worthwhile to compare the two approaches. While
NRQCD can in principle be generalized
to any order in $v$, the Schr\"odinger-Pauli approach
is only valid up to order $v^2$. Also, we cannot treat heavy-light
systems.  In NRQCD the zero point energy can be fixed by measuring
the dispersion relation while in our approach only properties of
particles at rest can be studied. The clear advantage of the
method presented here
is that with one simulation only, we easily obtain all spectral
properties (including arbitrary excitations)
for any (sufficiently heavy) quark mass. From
a two-body Hamiltonian formulation of the problem,
the effect of individual terms on the
spectrum becomes immediately apparent, and a transparent understanding
of the anatomy of the underlying interaction mechanism is obtained.
The potentials are protected by the global $Z_3$
symmetry~\cite{schlicht} from finite size effects, contrary to
NRQCD wave functions and masses, such that we can determine the potentials
for the $r$-range required, even
for broad excited state wave functions, on relatively
small spatial lattice volumes. 

The article is organized as follows: in Sec.~\ref{sec2}, we
introduce the Hamiltonian and present definitions of the potentials
that are suitable for lattice evaluation.
Moreover, we include theoretical expectations on the form of the
potentials.
Sec.~\ref{sec3} contains simulation details and
lattice specific techniques
wherever they differ from our SU(2)
investigation~\cite{wachter}. The renormalization of lattice operators
and the matching procedure between the effective nonrelativistic
theory and QCD are discussed in Sec.~\ref{sec3b}.
The resulting SU(3) potentials
are presented in Sec.~\ref{sec4}.
Promising results on charmonium and
bottomonium spectra are obtained and
discussed in Sec.~\ref{sec5}, before we conclude.

\section{The heavy quark potential}
\label{sec2}
\subsection{Hamiltonian formulation of the meson binding problem}
In Ref.~\cite{wachter}, we restricted ourselves to an
evaluation of SD corrections to the static
potential. Since we are going to include the complete
${\mathcal O}(v^2)$ corrections
into the present study and aim to predict quarkonia properties,
we find it worthwhile to briefly sketch some details of the
derivation of the Hamiltonian. The SD and MD parts as well as
relativistic corrections to the central potential have been
derived during the eighties~\cite{eich-fein,gromes,BBMP}.
The matching problem between QCD and the
effective Hamiltonian has been sorted out
to one-loop order for the SD terms
recently~\cite{Chen} and we extend this
to the remaining corrections.

It is instructive to start at ${\mathcal O}(v^0)$,
before proceeding to the ${\mathcal O}(v^2)$ Hamiltonian.
To this order, the heavy quark
propagator $S(x,y)=Q(x)Q^*(y)$ of a quark with mass $m$ obeys the
evolution equation in an external gauge field
$A_{\mu}$\footnote{Everything is consistently
rotated to Euclidean space-time.},
\begin{equation}
-\partial_4 S(x,y)=\left(igA_4 + m -\frac{{\mathbf D}^2}{2m}\right)S(x,y),
\end{equation}
where $D_{\mu}$ denotes the covariant derivative.
To ${\mathcal O}(v^0)$, the solution to the initial value problem,
\begin{equation}
S(x,y)|_{x_4=y_4}=\delta^3({\mathbf x}-{\mathbf y}),
\end{equation}
is given by,
\begin{equation}
S(x,y)=U({\mathbf x};x_4,y_4)
\,{\mathcal T}\exp
\left[-\int_{x_4}^{y_4}\!dt\,\left(m+\frac{{\mathbf p}^2(t)}{2m}\right)\right]
\delta^3({\mathbf x}-{\mathbf y}),
\end{equation}
where ${\mathcal T}$ denotes the time ordering operator.
$U({\mathbf x};x_4,y_4)$ is the static propagator of a quark,
traveling
from the point $({\mathbf x},x_4)$ to $({\mathbf x},y_4)$,
and consists of the corresponding temporal Schwinger
line times the factor $\exp\left(-E_0\tau\right)$,
with $\tau=y_4-x_4$.
$E_0(\mu)$ represents the static quark self-energy that
diverges like $\mu/\ln\mu$ with the cut-off scale $\mu$.

By combining two static propagators
into a Wilson loop,
one can determine the potential $V_0(r)$ between two static sources,
separated by a distance $r$,
in the limit of large Euclidian times,
\begin{equation}
\langle W(r,\tau)\rangle\propto\exp\left(-V_0(r)\tau\right),
\quad \left(\tau\rightarrow\infty\right).
\end{equation}
Note that the potential contains the static quark self-energies.
In order to obtain the spectrum of mesonic heavy quark bound states,
the Schr\"odinger equation (in the c.m.\ frame),
\begin{equation}
H\psi_{nlm}({\mathbf r})=E_{nl}\psi_{nlm}({\mathbf r}),
\end{equation}
can be solved, where the Hamiltonian,
\begin{equation}
H=2m+\frac{p^2}{m}+V_0(r),
\end{equation}
is determined
from combining two heavy quark propagators with each other.

We wish to study relativistic corrections to the $v=0$
limit; after a Foldy-Wouthuysen-Tani transformation,
the Feynman propagator is
expanded in terms of the heavy quark velocity\footnote{
Formally, this procedure is equivalent to expanding the Dirac
equation in powers of $1/c$ where $c$ denotes the speed of
light. Note, that in some of the NRQCD literature our ${\mathcal
O}(v^2)$ corrections are
counted as ${\mathcal O}(v^4)$.}, $v$,
around the static solution, in
order to determine the propagator
$K(x,y)=q(x)q^{\dagger}(y)$.
To ${\mathcal O}(v^2)$ the propagator is given
by,
\begin{equation}
-\partial_4K(x,y)=\left(igA_4+m-\delta m(\mu,m)+
O(x)\right)K(x,y),\label{kern}
\end{equation}
with the well known terms,
\begin{eqnarray}
O&=&-\frac{{\mathbf D}^2}{2m}
+\sum_{i=1}^4 c_i(\mu,m)O_i(\mu;m),\\
O_1(\mu;m)&=&\frac{({\mathbf D}^2)^2}{8m^3},\\
O_2(\mu;m)&=&\frac{g_s(\mu)}{2m}{\mathbf \sigma\cdot B},\\
O_3(\mu;m)&=&-i\frac{g_s(\mu)}{8m^2}({\mathbf D}\cdot{\mathbf E}
-{\mathbf E}\cdot{\mathbf D}),\\
O_4(\mu;m)&=&\frac{g_s(\mu)}{8m^2}{\mathbf \sigma}\cdot
({\mathbf D}\times{\mathbf E}
-{\mathbf E}\times{\mathbf D}).
\end{eqnarray}
$E_i$ and $B_i$ are
color-electric and -magnetic field components.
The heavy quark two-spinor $q(x)$
consists of the large components of the original
Dirac four-spinor after the Foldy-Wouthuysen-Tani rotation.

Since we have truncated the
expansion at a fixed power of $v$, we have
lost renormalizability and
the ultraviolet behavior is changed with respect to
QCD\footnote{This fact gave rise to a discussion on a supposed
discrepancy
between the Eichten-Feinberg-Gromes results~\cite{eich-fein,gromes}
and perturbative expansions~\cite{radford,Pantaleone,kummer} in powers 
of the coupling,
$g$, where additional terms that depend
logarithmically on the mass occur after
regulating loop diagrams. These terms
are now understood to arise from changes in
the ordering of integrations, and the
underlying problem is resolved~\cite{Chen}.}.
The theory is only effective
and valid in the range of small gluon momenta $q\leq \mu$.
Whenever the $O_i$ are determined at a scale $\mu$ 
that differs from $m$, the
couplings $c_i(\mu,m)$ (that are unity at tree-level)
have to be adjusted by matching the effective
theory to QCD at this scale; this guarantees the condition
$c_i(m,m)=1$ to hold. The zero point energy
is shifted by $\delta m(\mu,m)=E_0(\mu)-E_0(m)$,
with respect to QCD, where the static quark self-energy $E_0(\mu)$
can be estimated from perturbation
theory~\cite{hill1,davies,morningstar}.
Due to this self-energy, the pole mass is
shifted in respect to $m-\delta m$ within the
propagator:
$m^{\mbox{\scriptsize pole}}=m-\delta m+E_0(\mu)=
m+E_0(m)$.

Note that the Hamiltonian which corresponds to $K$ is identical to
that of NRQCD to order $v^2$ (up to irrelevant terms
that are introduced to remove doublers and stabilize the
evolution of the propagator on a discrete lattice with spacing $a$).
In the case of NRQCD, the effective lattice theory is matched to 
continuum
QCD in one step, such that the
coefficients $c_i$ do not only depend on $m/\mu$ (or $ma$) but
also on the lattice coupling $g(a)$. We
start from an effective theory,
formulated in the continuum, such that the matching procedures
(QCD-effective theory and continuum-lattice) will
be treated in two separate steps.

In addition, the operator, acting on $K$ in Eq.~(\ref{kern}), 
has the same structure to order $1/m^2$
as the Lagrangian of heavy quark
effective theory (HQET).
Therefore, the matching coefficients can be taken from
Refs.~\cite{falk,dugan,luke}\footnote{In
Refs.~\cite{hill1,luke,hill2},
it has been
shown that the kinetic energy term $-{\mathbf D}^2/(2m)$ does not
undergo renormalization. Unlike in lattice NRQCD, where the quark mass
becomes multiplicatively renormalized~\cite{davies}, here
the mass does not enter as a dynamical variable of the simulation,
but rather as an expansion parameter.
The correction to the kinetic
energy, that contains the dimension seven operator $q^{\dagger}{\mathbf
D}^4q$, however, is
accompanied by a nontrivial coefficient $c_1(\mu,m)$. This coefficient as well
as the mixing matrix between $O_1(\mu;m)$ and lower dimensional
operators has not yet been determined.
For this reason, for the time being,
we assume $c_1\approx 1$.},
\begin{eqnarray}
\label{c_2}
c_2(\mu,m)&=&\left(\frac{\alpha_s(\mu)}{\alpha_s(m)}\right)^{-\frac{9}{25}},\\\label{c_3}
c_3(\mu,m)&=&6\left(\frac{\alpha_s(\mu)}{\alpha_s(m)}
\right)^{-\frac{8}{25}}-5,\\\label{c_4}
c_4(\mu,m)&=&2c_2(\mu,m)-1.
\end{eqnarray}

In order to evaluate masses of heavy quarkonia, we have to combine
a propagator of a quark $q_1$ of mass
$m_1$ with one of an antiquark $q_2$
of mass $m_2$.
In following the steps of Refs.~\cite{BBMP,Chen}, one can obtain
the nonrelativistic Schr\"odinger-Pauli
Hamiltonian\footnote{The derivation of this expression
from QCD is nontrivial~\cite{BBMP}.} (in the c.m.\
system, 
i.e.\ ${\mathbf p}={\mathbf p_1}=-{\mathbf p_2}$ and ${\mathbf L}=
{\mathbf L_1}={\mathbf L_2}$),
\begin{equation}
\label{ham}
H=\sum_{i=1}^2\left(m_i-\delta m_i
+\frac{p^2}{2m_i}-c_1(m_i)\frac{p^4}{8m_i^3}\right)
+V(r,{\mathbf p},{\mathbf L},{\mathbf S_1},{\mathbf S_2}),
\end{equation}
where the potential,
\begin{equation}
V(r,{\mathbf p},{\mathbf L},{\mathbf S_1},{\mathbf S_2})=
\overline{V}(r)+
V_{\mbox{\scriptsize sd}}(r,{\mathbf L},{\mathbf S_1},{\mathbf S_2})+
V_{\mbox{\scriptsize md}}(r,{\mathbf p}),\label{prp}
\end{equation}
consists of a central part, SD and
MD corrections.

Note that under renormalization group transformations
the spin-spin interaction term of the effective two-particle
Lagrangian [$\int\! d^4x\, (q_1^{\dagger}\sigma\cdot{\mathbf B}q_1)
(q_2^{\dagger}\sigma\cdot{\mathbf B}q_2)$]
undergoes mixing with two local dimension six color
singlet two-fermion terms that have to be included at order $1/m^2$ into
the effective heavy quark Lagrangian. 
This very fact gives rise to the radiative
correction term~\cite{Chen}\footnote{We have substituted
the factor $8\pi C_F\alpha_s(\mu)\delta^3(r)$
of the reference by the potential
$V_4(r)$, which is equivalent at this
order in $\alpha_s$.},
\begin{equation}
\label{dh}
\delta V_{\mbox{\scriptsize sd}}=
-\frac{1}{3m_1m_2}{\mathbf S_1\mathbf S_2}\frac{3}{4}
c_2(m_2,m_1)\left(1-c_2^2(\mu,m_2)\right)
V_4(r),
\end{equation}
within the SD potential below.

The complete
result on the potential
to order $v^2$ with one-loop matching coefficients turns out to be,
\begin{eqnarray}
\overline{V}(r)&=&V_0(r)+\sum_{i=1}^2\frac{1}{8m_i^2}
c_3(m_i)\left(\nabla^2V_0(r)+\nabla^2V_a^E(r)\right)\nonumber\\
&-&\sum_{i=1}^2\frac{1}{8m_i^2}c_2^2(m_i)\nabla^2V_a^B(r),
\label{cepo}
\end{eqnarray}
\begin{eqnarray}
V_{\mbox{\scriptsize sd}}(r,{\mathbf L},{\mathbf S_1},{\mathbf S_2})
&=&\left(\frac{{\mathbf S_1}}{m_1^2}
+ \frac{{\mathbf S_2}}{m_2^2}\right){\mathbf L}
\frac{(2c_+-1)V_0'(r)+2c_+V_1'(r)}{2r}\nonumber\\
&+&\frac{{\mathbf S_1} + {\mathbf S_2}}{m_1m_2}{\mathbf L}
\frac{c_+V_2'(r)}{r}\nonumber\\\label{sdpo}
&+&\frac{S_1^iS_2^j}{m_1m_2}c_2(m_1)c_2(m_2)R_{ij}V_3(r)
\\\nonumber
&+&\frac{\mathbf S_1S_2}{3m_1m_2}
\left(c_2(m_1)c_2(m_2)-\frac{3}{4}c_2(m_2,m_1)\left[1-c_2^2(m_2)\right]
\right)V_4(r)\\
&+&\left(\frac{{\mathbf S_1}}{m_1^2}
- \frac{{\mathbf S_2}}{m_2^2}\right){\mathbf L}
\frac{c_-[V_0'(r)+V_1'(r)]}{r}\nonumber\\\nonumber
&+&\frac{{\mathbf S_1} - {\mathbf S_2}}{m_1m_2}{\mathbf L}
\frac{c_-V_2'(r)}{r}
\end{eqnarray}
and
\begin{eqnarray}
V_{\mbox{\scriptsize md}}(r,{\mathbf p})
&=&-\frac{1}{m_1m_2}
\left\{p_i,p_j,[\delta_{ij}V_b(r)-R_{ij}V_c(r)]\right\}_{\mbox{\scriptsize
Weyl}}\label{mdpo}\\
&+&\sum_{k=1}^2\frac{1}{m_k^2}\left\{p_i,p_j,[\delta_{ij}V_d(r)-R_{ij}V_e(r)]
\right\}_{\mbox{\scriptsize Weyl}}\nonumber
\end{eqnarray}
with
\begin{equation}
R_{ij}=\frac{r_ir_j}{r^2}-\frac{\delta_{ij}}{3},
\end{equation}
\begin{equation}
c_{\pm}=c_{\pm}(\mu,m_1,m_2)=\frac{1}{2}\left[c_2(\mu,m_1)\pm
c_2(\mu,m_2)\right],\quad m_1\geq m_2,
\end{equation}
\begin{equation}
c_i(m)=c_i(\mu,m).
\end{equation}
The symbol $\{a,b,c\}_{\mbox{\scriptsize Weyl}}=
\frac{1}{4}\{a,\{b,c\}\}$
denotes Weyl ordering of the three arguments.
$V_1',\ldots, V_4$ are related to spin-orbit and spin-spin interactions.
The MD potential
gives rise to correction terms of the
form $\frac{1}{r}L^2$, $\frac{1}{r^3}L^2$,
$\frac{1}{r}p^2$, $\frac{1}{r}$ and $\delta^3(r)$.
The correction to the static potential includes, besides $\nabla^2
V_a^E$ and $\nabla^2 V_a^B$, the expected Darwin term
$\nabla^2V_0$. 
Note that 
$V_1',\ldots, V_4$ as well as $\nabla^2V_a^E$
and $\nabla^2V_a^B$ depend on the matching scale $\mu$ while $V_0$
as well as
$V_b,\ldots, V_e$ are scale independent\footnote{
The latter potentials originate from 
perturbing a quark world line, along which the
field $A_4$ of Eq.~(\ref{kern})
contributes to the propagator, around the classical particle
trajectory. Since an overall renormalization of the gluon fields can
be absorbed into the quark wave function normalization, $V_b,\ldots, V_e$
are scale independent (like $V_0$).}.
In what follows, we will refer to the functions $V_1',\ldots, V_4$ as
SD potentials, $V_b,\ldots, V_e$ as MD potentials,
and $\nabla^2V_a^E$ and $\nabla^2V_a^B$ as corrections to the central
potential.

In order to derive the Hamiltonian from one-particle propagators, one
has to assume that interactions 
between the
two quarks are functions of a single global time coordinate
(instantaneous approximation).
Unlike NRQCD, the above Hamiltonian
cannot be generalized to higher orders in $v$ since this
would involve higher than first order temporal
derivatives of the quark momenta, which, on the quantum level,
cannot be reexpressed in terms of the canonical coordinates.

$V_0$, $\nabla^2V_a^E$, $\nabla^2V_a^B$, $V_1',\ldots, V_4$ and $V_b,\ldots,
V_e$ can be computed from lattice correlation functions
(in Euclidean time) of Wilson loop like operators.
Due to Lorentz invariance, certain pairs of potentials are related
to the static
potential by the Gromes~\cite{grome2} and BBMP~\cite{BBMP} relations,
\begin{eqnarray}
V_2'(\mu;r)-V_1'(\mu;r)&=&V_0'(r),\label{grom}\\
V_b(r)+2V_d(r)&=&\frac{r}{6}V_0'(r)-\frac{1}{2}V_0(r),\label{bram2}\\
V_c(r)+2V_e(r)&=&-\frac{r}{2}V_0'(r)\label{bram},
\end{eqnarray}
such that three potentials, e.g.\ $V_1'$, $V_d$ and $V_e$ can be
eliminated from the Hamiltonian. From arguments, similar to those
of Ref.~\cite{BBMP}, it is evident that the combination
$V_4(\mu;r)+2\nabla^2V_a^B(\mu;r)$ is a function of the static
potential and thus scale independent.
Given this observation, the structure of the Hamiltonian 
[Eqs.~(\ref{ham})--(\ref{sdpo})] and the Gromes
relation [Eq.~(\ref{grom})],
we can deduce the following one-loop
relations between potentials, evaluated at cut-off
scales $\mu_1$ and $\mu_2$,
\begin{eqnarray}
\nabla^2V_a^E(\mu_2;r)&=&c_3(\mu_1,\mu_2)\nabla^2V_a^E(\mu_1;r)
+\left[c_3(\mu_1,\mu_2)-1\right]\nabla^2V_0(r)\nonumber\\\label{vaesca}
&+&\left[1-c_2^2(\mu_1,\mu_2)\right]\left[\nabla^2V_a^B(\mu_1;r)
+\frac{7}{8}V_4(\mu_1;r)\right],\\
\nabla^2V_a^B(\mu_2;r)&=&\nabla^2V_a^B(\mu_1;r)
+\frac{7}{8}\left[1-c_2^2(\mu_1,\mu_2)\right]V_4(\mu_1;r),
\label{vabsca}\\
V_1'(\mu_2;r)&=&V_1'(\mu_1;r)-
\left[1-c_2(\mu_1,\mu_2)\right]V_2'(\mu_1;r),\label{v1sca}\\
V_2'(\mu_2;r)&=&c_2(\mu_1,\mu_2)V_2'(\mu_1;r),\label{v2sca}\\
V_3(\mu_2;r)&=&c_2^2(\mu_1,\mu_2)V_3(\mu_1;r),\label{v3sca}\\\label{v4sca}
V_4(\mu_2;r)&=&\frac{1}{4}\left[7c_2^2(\mu_1,\mu_2)-3\right]V_4(\mu_1;r).
\end{eqnarray}

\subsection{How to compute the potentials}
In the Schr\"odinger-Pauli approach, introduced above, the quarks interact
through a potential that only
depends on the distance, spins, and momenta of the sources,
Eqs.~(\ref{prp}) -- (\ref{mdpo}).
The time dependence has been separated and is implicitly included 
into coefficient functions of various interaction terms,
the central, SD and MD potentials.
These can be computed by a nonperturbative integration
over gluonic interactions. Therefore, the
potentials incorporate a summation over all possible
interaction times, $t$. One obtains the following expressions
in terms of expectation values in presence of a gauge
field background for the corrections to the static
potential~\cite{BBMP}\footnote{We have recast all
expressions
into forms that are more suitable for lattice simulations.
Via spectral decompositions of the underlying correlation
functions, equality between our definitions and those
of Refs.~\cite{eich-fein,gromes,BBMP}
can easily be shown.},
\begin{eqnarray}
\label{ce_1}
\nabla^2\hat{V}_a^E({\mathbf R})&=&2\lim\limits_{\tau\to\infty}
\int_0^\tau 
\!dt\, \langle\langle {\mathbf \hat{E}}({\mathbf 0},0)
{\mathbf \hat{E}}({\mathbf 0},t)\rangle\rangle_W^c,\\
\label{ce_2}
\nabla^2\hat{V}_a^B({\mathbf R})&=&2\lim\limits_{\tau\to\infty}
\int_0^\tau 
\!dt\, \langle\langle {\mathbf \hat{B}}({\mathbf 0},0)
{\mathbf \hat{B}}({\mathbf 0},t)\rangle\rangle_W,
\end{eqnarray}
where the superscript ``$c$'' denotes the connected part,
\begin{eqnarray}
\nonumber
\langle\langle \hat{E}_i({\mathbf n}_1,0)
\hat{E}_j({\mathbf n}_2,t)\rangle\rangle_W^c
&=&
\langle\langle \hat{E}_i({\mathbf n}_1,0)
\hat{E}_j({\mathbf n}_2,t)\rangle\rangle_W\\
&-&
\lim\limits_{t'\to\infty}\langle\langle \hat{E}_i({\mathbf n}_1,0)
\hat{E}_j({\mathbf n}_2,t')\rangle\rangle_W\label{disco}.
\end{eqnarray}
For the SD potentials one finds~\cite{eich-fein,gromes},
\begin{eqnarray}
\label{ef_1}
\frac{R_k}{R}\tilde{V}_1'({\mathbf R}) &=& 
2\epsilon_{ijk}
\!\lim\limits_{\tau\to\infty}\!
\int_0^\tau 
\!\!dt\,t \langle\langle \hat{B}_i({\mathbf 0},0)\hat{E}_j({\mathbf
0},t)
\rangle\rangle_W, \\
\label{ef_2}
\frac{R_k}{R}\tilde{V}_2'({\mathbf R}) &=& \epsilon_{ijk}
\!\lim\limits_{\tau\to\infty}\!
\int_0^\tau
\!dt\, t\langle\langle \hat{B}_i({\mathbf
0},0)\hat{E}_j({\mathbf R},t)
\rangle\rangle_W, \\
\label{ef_3}
R_{ij}\tilde{V}_3({\mathbf R}) &=& 2\lim\limits_{\tau\to\infty}
\int_0^\tau 
\!dt\,\left[\langle\langle \hat{B}_i({\mathbf 0},0)
\hat{B}_j({\mathbf R},t)\rangle\rangle_W\right.\\\nonumber
&-&\frac{\delta_{ij}}{3}\left.
\langle\langle {\mathbf \hat{B}}({\mathbf 0},0){\mathbf
\hat{B}}({\mathbf R},t)
\rangle\rangle_W\right],\\
\label{ef_4}
\tilde{V}_4({\mathbf R}) &=& 2\lim\limits_{\tau\to\infty}
\int_0^\tau 
\!dt\, \langle\langle {\mathbf \hat{B}}({\mathbf 0},0)
{\mathbf \hat{B}}({\mathbf R},t)\rangle\rangle_W.
\end{eqnarray}
Finally, the MD potentials are~\cite{BBMP},
\begin{eqnarray}
\label{md_1}
\hat{V}_b({\mathbf R})&=&-\frac{1}{3}\lim\limits_{\tau\to\infty}
\int_0^\tau 
\!dt\,t^2\langle\langle {\mathbf \hat{E}}({\mathbf 0},0)
{\mathbf \hat{E}}({\mathbf R},t)\rangle\rangle_W^c,\\
\label{md_2}
R_{ij}\hat{V}_c({\mathbf R}) &=& \lim\limits_{\tau\to\infty}
\int_0^\tau 
\!dt\,t^2\left[\langle\langle \hat{E}_i({\mathbf 0},0)
\hat{E}_j({\mathbf R},t)\rangle\rangle_W^c\right.\\\nonumber
&-&\frac{\delta_{ij}}{3}\left.
\langle\langle {\mathbf \hat{E}}({\mathbf 0},0){\mathbf
\hat{E}}({\mathbf R},t)
\rangle\rangle_W^c\right],\\
\label{md_3}
\hat{V}_d({\mathbf R})&=&\frac{1}{6}\lim\limits_{\tau\to\infty}
\int_0^\tau 
\!dt\,t^2\langle\langle {\mathbf \hat{E}}({\mathbf 0},0)
{\mathbf \hat{E}}({\mathbf 0},t)\rangle\rangle_W^c,\\
R_{ij}\hat{V}_e({\mathbf R}) &=&-\frac{1}{2}\lim\limits_{\tau\to\infty}
\int_0^\tau
\!dt\,t^2\left[\langle\langle \hat{E}_i({\mathbf 0},0)
\hat{E}_j({\mathbf 0},t)\rangle\rangle_W^c\right.\label{md_4}\\\nonumber
&-&\frac{\delta_{ij}}{3}\left.
\langle\langle {\mathbf \hat{E}}({\mathbf 0},0){\mathbf
\hat{E}}({\mathbf 0},t)
\rangle\rangle_W^c\right].
\end{eqnarray}
${\mathbf R}$ denotes a lattice vector of length $R=ra^{-1}$.
At small lattice spacing
$a$, the above potentials should approach their continuum
counterparts and rotational invariance is expected to be restored,
$\hat{V}_0({\mathbf R})=aV_0(r)$,
$\tilde{V}_{1,2}'({\mathbf R})=a^{2}V_{1,2}'(\mu;r)$,
$\tilde{V}_{3,4}({\mathbf R})=a^{3}V_{3,4}(\mu;r)$,
$\nabla^2\hat{V}_a^{E,B}({\mathbf R})=a^{3}\nabla^2V_a^{E,B}(\mu;r)$,
$\hat{V}_{b,c,d,e}({\mathbf R})=aV_{b,c,d,e}(r)$,
where $\mu=\pi/a$.

Throughout the previous equations,
the expectation value $\langle\langle F_1F_2\rangle\rangle_W$ is 
defined as,
\begin{equation}
\label{dex}
\left\langle\langle F_1F_2\rangle\right\rangle_W=
\frac{\langle\mbox{Tr}\,{\mathcal P}
\left[\exp\left(ig\int_{\partial W} dx_\mu\,A_\mu\right)F_1F_2\right]\rangle}{
\langle\mbox{Tr}\,{\mathcal P}
\left[\exp\left(ig\int_{\partial W} dx_\mu\,A_\mu\right)\right]\rangle},
\end{equation}
where $\partial W$ represents a closed path [the contour of a Wilson loop
$W({\mathbf R},T)$] and
${\mathcal P}$ denotes path ordering of the arguments.
Although we have chosen a lattice inspired notation for the potentials
[Eqs.~(\ref{ce_1})--(\ref{md_4})], so far everything is generally
applicable to lattice as well as continuum formulations of QCD.
In following Huntley and Michael (HM)~\cite{huntley}, we implement the
discretized version
of Eq.~(\ref{dex}),
\begin{equation}
\label{defcor}
\langle\langle
\hat{F}_1\hat{F}_2\rangle\rangle_W
=-\frac{\left\langle
{\mathcal P}\left[W\,(\Pi_1-\Pi^{\dagger}_1)_{\mbox{\scriptsize tl}}
(\Pi_2-\Pi^{\dagger}_2)_{\mbox{\scriptsize tl}}\right]
\right\rangle\left\langle\left. W\right.\right\rangle}
{\left\langle {\mathcal
P}\left[W(\Pi_1+\Pi^{\dagger}_1)\right]\right\rangle
\left\langle {\mathcal
P}\left[W(\Pi_2+\Pi^{\dagger}_2)\right]\right\rangle},
\end{equation}
where the subscript $i=1,2$
represents the multi-index $(n_i,\mu_i,\nu_i)$ and
$n_i$ are integer valued
four-vectors. The subscript ``tl'' indicates that only
the traceless part is to be taken:
$(A)_{\mbox{\scriptsize tl}}=A-\frac{1}{3}\mbox{Tr}A$.
$\hat{F}_i$ are related to the electric and magnetic
fields in the following way,
\begin{equation}
\hat{F}_{\mu\nu}=ga^2F_{\mu\nu},\quad
\hat{E}_i= \hat{F}_{i4},\quad
\hat{B}_i=\frac{1}{2}\epsilon_{ijk}\hat{F}_{jk}.
\end{equation}
These conventions eliminate imaginary phases and
factors $g^2a^4$ from Eqs.~(\ref{ce_1})--(\ref{md_4}).

We have taken $\Pi_{\mu\nu}(n)$ to
be the spatial average of the
four (two) plaquettes, enclosing the lattice point
$n$ for magnetic (electric) fields,
\begin{equation}\label{epi}
\Pi_{ij}(n) = \frac{1}{4}\left[P_{i,j}(n)+ P_{i,-j}(n)
+P_{-i,-j}(n)+P_{-i,j}(n)\right]
\end{equation}
and
\begin{equation}\label{epie}
\Pi_{i4}\left(n+\frac{1}{2}\hat{4}\right)=
\frac{1}{2}\left[P_{i,4}(n)+P_{-i,4}(n)\right]
\end{equation}
with
\begin{equation}
P_{\mu,\nu}(n)=U_{\mu}(n)U_{\nu}(n+\hat{\mu})U^{\dagger}_{\mu}(n+\hat{\nu})
U^{\dagger}_{\nu}(n),\quad
U_{-\mu}(n)=U^{\dagger}_{\mu}(n-\hat{\mu}).
\end{equation}
With this choice of $\Pi$, Eq.~(\ref{defcor}) is correct up to
${\mathcal O}(a^2)$, the discretization error of the Wilson action, used
for generating the gauge field background. 
Note that the electric
fields are living at half-integer time coordinates, in between two
adjacent spatial lattice hyperplanes.
$U_{\mu}(n)$ is the $SU(3)$
link variable, related to the field
$A_{\mu}(x)$ at $x=(n+\frac{1}{2}\hat{\mu})a$:
$U_{\mu}(n)={\mathcal P}\exp
\left[iga\int_{n}^{n+\hat{\mu}}dn'_{\nu}A_{\nu}(n'a)\right]$.

In practical computation, the temporal extent $T$ of the Wilson loop
$W$ [within Eqs.~(\ref{ce_1})--(\ref{md_4})] is 
adapted according to the
formula $T=t+\Delta t_1+\Delta t_2$. $\Delta t_i$,
the separations of the ``ears'' $F_1$ and $F_2$ from the corresponding
spatial closures of the
Wilson loop, are kept fixed throughout the simulation while
the interaction time $t$, is varied. The discretized version of the
nominator of the correlation function
Eq.~(\ref{dex}) is visualized for the case of an electric
and a magnetic ear in Fig.~\ref{ohr}.
Strictly speaking, 
Eqs.~(\ref{ce_1})--(\ref{md_4}) apply in the limits $\Delta
t_i\rightarrow\infty$ only. 
$\Delta t_1$ ($\Delta t_2$) represents
the time the gluon field has to
decay into the ground-state, after (before)
creation (annihilation) of the $qq^{\dagger}$ state and is a
control parameter of the simulation.

\subsection{Theoretical expectations}
\subsubsection{General considerations}
In addition to the exact Gromes and BBMP
constraints, Eqs.~(\ref{grom})--(\ref{bram}),
some approximate relations between the SD
potentials are anticipated from exchange symmetry arguments.
We start from the standard assumption that the origin of the
static potential is due to vector- and scalar-like gluon exchange
contributions.
Given that a vector-like exchange can
grow at most logarithmically with $r$~\cite{grome}, the nature of the
linear part of the
confining potential can only be scalar. As we will see, $V_2'(r)$ is
short ranged, such that the confining part only contributes to
$V_1'(r)$. This leads us to expect $V_2'(r)$ to be purely vector-like.
Under the additional assumptions that pseudoscalar contributions can
be neglected and that $V_1'$ does not contain a vector-like
piece, one ends up with the scenario of interrelations~\cite{grome2},
\begin{eqnarray}
\label{exv31}
V_3(r)&=&\frac{V_2'(r)}{r}-V_2''(r),\\\label{exv3}
V_4(r)&=&2\nabla^2V_2(r),
\end{eqnarray}
which of course has to 
be in agreement with leading order perturbation
theory. However, Eqs.~(\ref{exv31})--(\ref{exv3}) hold true for any
effective gluon propagator that transforms like a Lorentz vector.
Unlike the Gromes relation,
the above relations cannot
be exact, which is evident from
Eqs.~(\ref{v2sca})--(\ref{v4sca}).

\subsubsection{One gluon exchange potentials}
\label{expect}
In order to parameterize the short range behavior of the
potentials, it is useful to resort to weak coupling perturbation
theory. For modeling of lattice artefacts, we have
calculated the SD potentials
to tree-level in Ref.~\cite{wachter}.
Here, we supplement these results by the remaining ${\mathcal O}(v^2)$
potentials.

{\bf Lattice potentials.}
In the following we will use the conventions,
\begin{equation}
\label{a10}
G_L({\mathbf R})=\frac{1}{L_{\sigma}^3}
\sum_{{\mathbf q}\ne {\mathbf 0}}\frac{e^{i{\mathbf q}{\mathbf R}}}
{\sum_i\hat{q}_i^2},\quad\hat{q}_i=2\sin\left(\frac{q_i}{2}\right)
\end{equation}
and
\begin{equation}
F_L({\mathbf R})=\frac{2}{L_{\sigma}^3}
\sum_{{\mathbf q}\ne {\mathbf 0}}\frac{e^{i{\mathbf q}{\mathbf R}}}
{\left(\sum_i\hat{q}_i^2\right)^2},
\end{equation}
with
\begin{equation}
q_i= \frac{2\pi}{L_{\sigma}} m_i, \quad
m_i=-\frac{L_{\sigma}}{2}+1,\ldots,\frac{L_{\sigma}}{2}.
\end{equation}
$L_{\sigma}$ denotes the number of 
lattice sites along a linear spatial extent.
Note that the above functions have the large-$R$
behavior (for $L_{\sigma}\gg R$),
\begin{equation}
G_L({\mathbf R})\rightarrow\frac{1}{4\pi R},\quad
F_L({\mathbf R})\rightarrow F_L({\mathbf 0})-\frac{1}{4\pi}R,\label{propag}
\end{equation}
where $F_L({\mathbf 0})$ diverges linearly with $L_{\sigma}$.

We find
\begin{eqnarray}\label{aa6}
\hat{V}_0({\mathbf R})&=&-C_Fg^2\left(G_L({\mathbf R})-G_L({\mathbf
0})\right),\\
\label{v2p}
\tilde{V}_2'({\mathbf R})&=&-\frac{R}{R_j}C_Fg^2\Delta_j\Xi_j^{(\perp)}
G_L({\mathbf R}),\\
\label{latv31}
\tilde{V}_3({\mathbf R})&=&
\frac{R^2}{R_jR_k}
C_Fg^2\Delta_j\Delta_k\Xi_iG_L({\mathbf R}),
\end{eqnarray}
for $j\ne k$, $i\ne j$, $i\ne k$ and $R_j\ne 0$, $R_k\ne 0$.
Unless explicitly stated, no summations over indices that appear
twice are performed.
For $2R_i^2\ne R_j^2+R_k^2$, $i, j, k$ as above, we can derive
the expression,
\begin{equation}
\tilde{V}_3
=\frac{R^2}{2R_i^2-R_j^2-R_k^2}
C_Fg^2\frac{1}{2}\left[4\Delta_i^{(2)}\Xi-\Delta_j^{(2)}(\Xi_k+\Xi)
-\Delta_k^{(2)}(\Xi_j+\Xi)\right]
G_L({\mathbf R}).\label{latv32}
\end{equation}
The remaining potentials are given by,
\begin{eqnarray}\label{latv4}
\tilde{V}_4({\mathbf R})
&=&-2C_Fg^2\sum_i\Delta^{(2)}_i\Xi^{(\perp)}_i\,
G_L({\mathbf R}),\\\label{latvb}
\hat{V}_b({\mathbf R})&=&\frac{C_Fg^2}{6}\left(\sum_i\Delta_i^{(2)}\Xi_i\,
F_L({\mathbf R})+6\Xi G_L({\mathbf R})\right),\\\label{latvc}
\hat{V}_c({\mathbf R})&=&\frac{C_Fg^2}{2}\frac{R^2}{3R_iR_j-\delta_{ij}R^2}
\left[-3\Delta_i\Delta_jF_L({\mathbf R})\right.\nonumber\\
&+&\left.\delta_{ij}
\left(\sum_k\Delta_k^{(2)}\Xi_k\, F_L({\mathbf
R})+6\left(\Xi-\Xi_i\right)G_L({\mathbf
R})\right)\right],\label{abc6}\\
\hat{V}_d({\mathbf R})&=&-\frac{C_Fg^2}{4}\left[G_L({\mathbf
0})+G_L(1)+\frac{1}{2}F_L(2)
-\frac{1}{2}F_L({\mathbf 0})\right]
\nonumber\\
&=&-\frac{C_Fg^2}{4}G_L({\mathbf 0})\quad
(L_{\sigma}\rightarrow\infty),\label{vdpt}\\
\nabla^2\hat{V}_a^E({\mathbf R})&=&-3C_Fg^2\left[G_L({\mathbf
0})-G_L(2)\right]\approx -0.629525 C_Fg^2,\label{abd6}\\
\nabla^2\hat{V}_a^B({\mathbf R})&=&-6C_Fg^2\left[G_L({\mathbf
0})-G_L(\sqrt{2})\right]\approx -1.185237 C_Fg^2,\label{abe6}
\end{eqnarray}
with $f(1)=f({\mathbf \hat{1}})$, $f(\sqrt{2})=f({\mathbf
\hat{1}}+{\mathbf \hat{2}})$, $f(2)=f(2{\mathbf \hat{1}})$. The numerical
values refer to the infinite volume limit. In this limit, one obtains
$G_L({\mathbf 0})\approx 0.2527310$.
The potentials
$\tilde{V}_1'$ and $\hat{V}_e$ vanish to lowest order
perturbation theory. 

The Casimir factor of SU(3) gauge theory is $C_F=4/3$.
The $\Delta$s and $\Xi$s denote the 
finite difference and averaging operators,
\begin{eqnarray}
\Delta_i f({\mathbf n})&=&\frac{1}{2}
\left[f({\mathbf n}+{\mathbf \hat{i}})
-f({\mathbf n}-{\mathbf \hat{i}})\right],\\
\Delta_i^{(2)}f({\mathbf n})&=&f({\mathbf n}+{\mathbf \hat{i}})
-2f({\mathbf n})
+f({\mathbf n}-{\mathbf \hat{i}}),\\
\Delta^{(2)}&=&\sum_i \Delta_i^{(2)},\\
\Xi_i f({\mathbf n})&=&\frac{1}{4}\left[
f({\mathbf n}+{\mathbf \hat{i}})
+2f({\mathbf n})
+f({\mathbf n}-{\mathbf \hat{i}})\right],\\
\Xi_i^{(\perp)}&=&\frac{1}{2}\sum_{j\ne i}\Xi_j,\quad
\Xi=\frac{1}{3}\sum_{i}\Xi_i.
\end{eqnarray}

In Ref.~\cite{wachter}, we have proven that an exact lattice analogue
to the Gromes relation does not exist. However,
the Gromes as well as the BBMP relations will be retrieved in the continuum
limit and approximately hold within the scaling region on the lattice
for $R\gg 1$.

{\bf Continuum potentials.}
In continuum perturbation theory, one
obtains the following tree-level expressions,
\begin{eqnarray}
V_0(r)&=&-C_F\alpha_s\int\frac{dq^3}{2\pi^2}\frac{e^{i{\mathbf
qr}}}{q^2}
=-C_F\frac{2\alpha_s}{\pi}\int_0^{\infty}\!dq\,\frac{\sin qr}{qr}\label{a111}\\
V_2'(r)&=&-iC_F\alpha_s\int\frac{dq^3}{2\pi^2}\frac{{\mathbf qr}}{q^2r}
e^{i{\mathbf qr}}
=-C_F\frac{2\alpha_s}{\pi}\int_0^{\infty}\!dq\,q^2r\, j_1(qr)
\label{a222}\\
V_3(r)&=&-C_F\alpha_s\int\frac{dq^3}{2\pi^2} \frac{({\mathbf
qr})^2}{q^2r^2}e^{i{\mathbf qr}}
=-C_F\frac{2\alpha_s}{\pi}\int_0^{\infty}\!dq\,q^2j_2(qr)\label{a333}\\
V_4(r)&=&C_F\alpha_s\int\frac{dq^3}{2\pi^2} e^{i{\mathbf
qr}}=C_F\frac{2\alpha_s}{\pi}\int_0^{\infty}\!dq\,q^2\frac{\sin
qr}{qr},\label{a444}
\end{eqnarray}
with $\alpha_s = g^2/(4\pi)$.
$V_b$ and $V_c$ are given by $-2V_0/3$ and $V_0/2$, respectively.
The self-energy $C_V=a^{-1}C_Fg^2G_L({\mathbf 0})$ has been subtracted
from $V_0$.
$\tilde{V}_1'$ and $\hat{V}_e$ vanish to lowest order
perturbation theory while $\nabla^2\hat{V}_a^E$, $\nabla^2\hat{V}_a^B$
and $\hat{V}_d$ only contain diverging self-energy contributions.
A linear confining contribution can be
introduced by adding a $-1/q^4$-term to
$V_0$ in momentum space.
The integrals for the SD potentials are
suppressed in the infrared region like $q^2$ or $q^3$,
such that we naively expect perturbation
theory to be more reliable in this case than for the static
potential or $V_b$ and $V_c$.

Eqs.~(\ref{a111})--(\ref{a444}) yield,
\begin{eqnarray}
V_0(r)&=&- C_F\frac{\alpha_s}{r},\label{vp0}\\
V_2'(r)&=& C_F\frac{\alpha_s}{r^2},\label{vp2}\\
V_3(r)&=& 3C_F\frac{\alpha_s}{r^3},\label{vp3}\\
V_4(r)&=& 8\pi C_F\alpha_s \delta^3(r),\label{vp4}\\
V_b(r)&=& \frac{2}{3}C_F\frac{\alpha_s}{r},\label{vpb}\\
V_c(r)&=&-\frac{1}{2}C_F\frac{\alpha_s}{r},\label{vpc}
\end{eqnarray}
in agreement with the large-$R$ (i.e.\ $r\gg a$)
expectations of Eqs.~(\ref{aa6})--(\ref{abc6}) [cf.\ Eq.~(\ref{propag})].

\subsubsection{Large distance behavior}
In combining the large-$r$ behavior from the
minimal area law (MAL) of fluctuating
world sheets~\cite{BBMP} (including a perimeter term)
with the expectation from tree-level perturbation theory,
Eqs.~(\ref{vp0})--(\ref{vpc}),
one obtains,
\begin{eqnarray}
V_0(r)&=&V_c-\frac{e}{r}+\kappa r,\label{v0pa}\\
\nabla^2V_a^E(r)&=&C_a^E-\frac{b}{r},\quad\nabla^2V_a^B(r)=C_a^B\label{vapa}\\
V_1'(r)&=&-\frac{h}{r^2}-\kappa,\quad V_2'(r)=\frac{e-h}{r^2},\label{v1pa}\\
V_3(r)&=&3\frac{e-h}{r^3},\quad V_4(r)=8\pi (e-h)\delta^3(r),\label{v3pa}\\
V_b(r)&=&C_b+\frac{2}{3}\frac{e}{r}-\frac{\kappa}{9}r,\quad
V_c(r)=-\frac{1}{2}\frac{e}{r}-\frac{\kappa}{6}r,\label{vbpa}\\
V_d(r)&=&C_d-\frac{\kappa}{9} r,\quad
V_e(r)=-\frac{\kappa}{6} r,\label{mdpa}
\end{eqnarray}
with $e=C_F\alpha_s$,
in agreement with the Gromes and BBMP relations,
Eqs.~(\ref{grom})--(\ref{bram}).
From the perimeter term in MAL
one obtains $C_a^E=C_a^B=C_b=0$ and $C_d=-C_V/4$.
In tree-level perturbation theory, one finds a consistent infinite volume
result $C_d=-C_V/4$
[Eq.~(\ref{vdpt})], while $C_a^E$
comes out to be significantly smaller than $C_a^B$
[Eqs.~(\ref{abd6})--(\ref{abe6})].
However, our numerical data shows $C_a^E\approx C_a^B$, in
agreement with the MAL result.
{}From Eqs.~(\ref{vaesca})--(\ref{v1sca}),
it is obvious that the tree-level perturbative expectations
Eqs.~(\ref{vp0})--(\ref{vpc}) cannot adequately describe the
potentials at all scales,
$\mu$. In general, $V_1'$ and $V_2'$ will undergo mixing, such that
$V_1'$ will attain a Coulomb-like contribution. For the same reason,
$\nabla^2V_a^E$ is expected to include a $1/r$-piece.
We have accounted for this fact by allowing for two 
additional parameters, $b$ and $h$. In principle, $\nabla^2V_a^{E,B}$
can also contain $\delta$-like admixtures [Eqs.~(\ref{vaesca}),
(\ref{vabsca})], which we have ignored in Eq.~(\ref{vapa}).

\section{Lattice simulations}
\label{sec3}
In Ref.~\cite{wachter}, we have developed suitable techniques
for a lattice evaluation of the potentials and applied them to
SU(2) gauge theory. We investigated possible sources of
systematic errors such as finite size effects. In this section,
we describe details of our SU(3) simulations, insofar these
differ from the SU(2) study.

\subsection{Simulation parameters}
\label{sec31}
We analyse
two sets of Monte Carlo configurations that have been generated
with the standard Wilson action on hypercubic
lattices of volumes $V=L_{\sigma}^3L_{\tau}=16^4$ at $\beta=6.0$ and
$V=32^4$ at $\beta=6.2$ (Table~\ref{et1}).
The above couplings correspond to inverse
lattice spacings $a^{-1}\approx 2.1$~GeV and
$a^{-1}\approx 2.9$~GeV, respectively. The scale has
been determined from the value $\sqrt{\kappa}=468$~MeV for the string
tension that we obtain from the fit to the bottomonium spectrum of
Sec.~\ref{sec5}.
The number of
independent Monte Carlo configurations $n_{\mbox{\scriptsize conf}}$,
generated at each set of parameters, is included into the table. 
Based on previous experience, we expect
finite size effects to be below statistical accuracy at these
volumes~\cite{pot,pot2,wachter}.
For the updating of the gauge fields, a hybrid of
Fabricius-Haan heatbath~\cite{fh} and an
overrelaxation algorithm has been implemented~\cite{hor}.
Within both procedures, we successively update the three diagonal $SU(2)$
subgroups of a given link~\cite{cabibbo}. The heatbath sweeps
have been randomly mixed with overrelaxation steps with probability
$1/7$. The links have been visited in lexicographical ordering within
hypercubes of $2^4$ lattice sites, i.e., within each such hypercube,
first all links pointing
into direction $\hat{1}$ are visited site by
site, then all links in direction $\hat{2}$ etc.. After 2000 initial
heatbath thermalization sweeps in either case, measurements are taken
every 100 sweeps to ensure de-correlation. We find no evidence for any
autocorrelation effects between these configurations.

\subsection{Noise reduction}
Statistical fluctuations have been reduced
by ``integrating out'' temporal links that appear within the Wilson
loops and the electric ears analytically, wherever possible.
By ``link integration'' we mean the following substitution~\cite{parisi},
\begin{equation}
U_4(n)\longrightarrow W_4(n)=\frac{1}{Z}\frac{\partial Z}{\partial
F^{\dagger}_{\mu}(n)}=
\frac{\int_{SU(3)}\!dU\,Ue^{S_{n,4}(U)}}
{\int_{SU(3)}\!dU\,e^{S_{n,4}(U)}},
\end{equation}
with
\begin{equation}
S_{n,\mu}(U)=\mbox{Tr}\left(F_{\mu}(n)U^{\dagger}+UF_{\mu}^{\dagger}(n)\right),
\end{equation}
and
\begin{equation}
F_{\mu}(n)=\frac{\beta}{6}
\sum_{\nu\neq\mu}U_{\nu}(n)U_{\mu}(n+\hat{\nu})
U_{\nu}^{\dagger}(n+\hat{\mu}),\quad\beta=\frac{6}{g^2}.
\end{equation}
$W_4(n)$ is in general no longer an $SU(3)$ element.
In this way, time-like links are replaced
by the mean value they take in the neighborhood of
the enclosing staples $F_4(n)$. 
Only those links that do not share a common plaquette can be
integrated independently, without changing expectation values.

We have attempted to compute $W_4(n)$ analytically for the case of
SU(3) gauge theory. Based on the character expansion of $SU(N)$
matrices of Ref.~\cite{brower}, the following expression can be
obtained\footnote{For simplicity, we suppress spatial coordinates and
Dirac indices of $U$, $F$ and $W$.},
\begin{eqnarray}
Z(F)&=&\int_{SU(3)}\!dU\,\exp\left(\mbox{Tr}\left(FU^{\dagger}
+UF^{\dagger}\right)\right)\\
&=&\oint\frac{dx}{2\pi i}\oint\frac{dy}{2\pi i}
\exp\left(xQ+y+\frac{P(x)}{xy}\right),
\end{eqnarray}
with
\begin{eqnarray}
Q &=& \det(F)+\det(F^{\dagger}),\\
P(x)&=&1+\mbox{Tr}(FF^{\dagger})x+ \frac{1}{2}\left[
\mbox{Tr}^2(FF^{\dagger})-\mbox{Tr}(FF^{\dagger})^2\right]x^2
+\det(FF^{\dagger})x^3.
\end{eqnarray}
{}From
\begin{eqnarray}
J_n&=&\oint\frac{dx}{2\pi i}x^n\frac{e^{Qx}}{R(x)}I_1(2R(x)),\label{oi1}\\
K_n[O]&=&\oint\frac{dx}{2\pi i}Ox^n\frac{e^{Qx}}{P(x)}I_2(2R(x)),\label{oi2}
\end{eqnarray}
where $I_n$ denote the modified Bessel functions and
$R(x)=\sqrt{P(x)/x}$,
one obtains~\cite{forcrand2},
\begin{equation}
Z=J_0,\quad \frac{\partial Z}{\partial F^{\dagger}}
=J_1\frac{\partial Q}{\partial F^{\dagger}}
+K_0\left[\frac{\partial P}{\partial F^{\dagger}}\right],
\end{equation}
such that
\begin{equation}
W=\frac{1}{J_0}\left[J_1G
 +K_1F+K_2F\left(\mbox{Tr}(F^{\dagger}F)
-F^{\dagger}F\right)+K_3\det(F)G\right],
\end{equation}
where $K_n=K_n[1]$,
$G_{il}=\frac{1}{2}\epsilon_{ijk}\epsilon_{lmn}F_{jm}^*F_{kn}^*$.
Note that $J_n$ and $K_n$ are real numbers.
For the computation of
Bessel functions we use the asymptotic expansion,
\begin{equation}
I_n(z)=\frac{e^z}{\sqrt{2\pi
z}}\sum_{j=0}^{\infty}(-1)^j\frac{A_j(n)}{z^j}
\end{equation}
with
\begin{equation}
A_j(n)=\frac{(4n^2-1^2)(4n^2-3^2)\cdots(4n^2-(2j-1)^2)}{8^jj!},
\end{equation}
up to fifth order in $j$.
The above expansion is valid for arguments,
$z=2R(x)$, with large modulus. A circular integration path
with radius $|x|=0.015$ turns out to be appropriate~\cite{phd}
at $\beta\approx 6$. 
A Gaussian quadrature algorithm with
64 abscissas is used.
By exploiting the symmetry of the contour integrals
Eqs.~(\ref{oi1})
and (\ref{oi2}) under the
transformation $x\rightarrow -x$, we are able to
reduce the computational effort
by a factor 2.

\subsection{Ground state enhancement}
In this section, we will discuss the control of excited state
contributions at finite de-excitation times $\Delta t_i$.
We found $\Delta t = 2$ to be appropriate for magnetic ears 
and $\Delta t=3/2$ 
for electric ears (see Fig.~\ref{ohr}).
The spatial transporters within the Wilson loops have been
smeared to suppress excited state pollutions from the very beginning.
Our smearing procedure~\cite{schlicht,APE}
consists of iteratively replacing each
spatial link
\(U_i(n)\) within the Wilson loop by a ``fat'' link,
\begin{equation}
\label{sme}
U_i(n)\rightarrow {\mathcal N}\left(\alpha\, U_i(n)+\sum_{j\neq
i}U_j(n)U_i(n+\hat{j})U_j^{\dagger}(n+\hat{i})\right),
\end{equation}
with free parameter $\alpha$. ${\mathcal N}$ denotes an
operator that projects the argument back into
the $SU(3)$ group: $U={\mathcal N}(A)\in SU(3)$
with $\mbox{Re Tr}\{A^{\dagger}U\}=\max$.
Within this procedure, the (spatial) links are visited in the same
lexicographical ordering as within the Monte Carlo updating of gauge
configurations. We find satisfactory ground-state enhancement with the
parameter choice
$n_{\mbox{\scriptsize iter}}=100$ and $\alpha=2$.

{}From expectation values of Wilson loops, the static
interquark potential can be determined in the limit of large $T$,
\begin{equation}
\langle W({\mathbf R},T)\rangle=C_0e^{-\hat{V}_0T}\left(1+\sum_{n>0}
\frac{C_n}{C_0}e^{-\Delta\hat{V}_nT}\right),
\end{equation}
where $\Delta\hat{V}_n=\hat{V}_n-\hat{V}_0$ denotes the gap between
the ground-state and the $n$th excited state (hybrid) potential.
The ${\mathbf R}$ dependency has been omitted from the
overlap coefficients $0\leq C_n\leq 1$ and potentials
$\hat{V}_n$, $\Delta\hat{V}_n$.
The smearing procedure results in an
increased weight $C_0$ (with respect to $C_n$, $n\geq 1$).
In Fig.~\ref{v0} the resulting static interquark potentials at
$\beta=6.0$ and $\beta=6.2$ are
displayed. All ground-state overlaps turn out to be well above 0.8.

Previous authors~\cite{rebbi,forcrand,koike,laermann}
have replaced the integrals
over interaction times by discrete sums.
This results in cut-off errors due to the 
finiteness of the integration bound $\tau$
[Eqs.~(\ref{ce_1})--(\ref{md_4})] as well
as ${\mathcal O}(a^2)$ integration errors. Both sources
of systematic uncertainties can be studied and
reduced by exploiting transfer matrix techniques.
In the following, we will
briefly summarize some of the results we have already presented in
Ref.~\cite{wachter}. 
The ratio of correlation functions between eared Wilson loops,
Eq.~(\ref{defcor}), is given by\footnote{The formula
applies to the case $\Delta t=\Delta t_1=\Delta t_2$. It
remains valid on a qualitative level
for $\Delta t_1\neq\Delta t_2$ with
$\Delta t =\min\{\Delta t_1,\Delta t_2\}$.},
\begin{equation}\label{eq:correl}
\langle\langle
\hat{F}_1\hat{F}_2\rangle\rangle_W=
\sum_m D^{12}_me^{-\Delta \hat{V}_mt}\left[1+E^{12}_me^{-\Delta
\hat{V}_1\Delta t}+\cdots \right].
\end{equation}
All constants are understood to depend on ${\mathbf R}$.
For details on $D^{12}_m$ and $E^{12}_m$, that
are functions of the spatial positions and the color-electric or -magnetic
components $F_1$ and $F_2$, see~\cite{wachter}.
The unwanted excited state contributions are suppressed by
factors $|E^{12}_m|\leq\sqrt{C_1/C_0}$ as well as
by $e^{-\Delta \hat{V}_1\Delta t}$.
The smallest value of $\Delta t$ that appears within an integral over
interaction times will determine the reliability of the
result. The bosonic string picture yields the large-$R$ expectation
$\Delta \hat{V}_1(R)= \pi/R$~\cite{luescher} for the lowest lying
hybrid
potential, which has been
qualitatively confirmed in numerical studies~\cite{huntley2,green}.

The cylindrically symmetric
creation operator that we use only projects onto states
within the $A_{1g}$ representation of the appropriate symmetry group
$D_{4h}$~\cite{d4h}. The lowest continuum angular momentum to
which it couples is $L=0$. The hybrid ($L=1$) state $E_u$ is the next
excitation~\cite{huntley2}. The operators used as magnetic ears
have no overlap with the $A_{1g}$ state, such that all correlation
functions that involve a magnetic ear
decay exponentially with Euclidean time.
This does not hold true for some of the correlators within 
the MD potentials and $\nabla^2\hat{V}_a^E$;
those electric ears which are not orthogonal to
${\mathbf R}$ have a nonvanishing
overlap with $A_{1g}$, such that the
disconnected part in Eq.~(\ref{eq:correl}),
$D_0^{12}$, does not vanish and has to be
explicitly subtracted in Eqs.~(\ref{ce_1}) and ({\ref{md_1})--(\ref{md_4}).

Note that
the $D_m^{12}$ are not normalized and can be negative. However, due to
invariance 
under time inversion, the correlation functions for $\tilde{V}_1'$ and
$\tilde{V}_2'$
[Eqs.~(\ref{ef_1} and (\ref{ef_2})] have to vanish at $t=0$, such that
$\sum_m D_m^{12}=0$ in this case.
In combining Eq.~(\ref{eq:correl}) with 
Eqs.~(\ref{ce_1})--(\ref{ce_2}) or (\ref{ef_3})--(\ref{ef_4}),
we obtain,
\begin{equation}
\nabla^2\hat{V}_a^{E,B},\tilde{V}_{3,4}
\propto \sum_{m>0}\int_0^{\infty}\!dt\,D_m^{12}
e^{-\Delta\hat{V}_mt}=
\sum_{m>0} \frac{D_m^{12}}{\Delta\hat{V}_m},
\end{equation}
with appropriate color field positions ${\mathbf
n}_1, {\mathbf n}_2$ and components
$\mu_1, \nu_1, \mu_2, \nu_2$.
Eqs.~(\ref{ef_1}) and (\ref{ef_2}) yield,
\begin{equation}
\tilde{V}_{1,2}' \propto \sum_{m>0}\int_0^{\infty}\!dt\,t\,D_m^{12}
e^{-\Delta\hat{V}_mt}
= \sum_{m>0} \frac{D_m^{12}}{\left(\Delta\hat{V}_m\right)^2},
\end{equation}
while from Eqs.~(\ref{md_1})--(\ref{md_4}) we obtain,
\begin{equation}
\hat{V}_{a,b,c,d} \propto \sum_{m>0}\int_0^{\infty}\!dt\,t^2\,D_m^{12}
e^{-\Delta\hat{V}_mt}
= \sum_{m>0} \frac{D_m^{12}}{\left(\Delta\hat{V}_m\right)^3}.
\end{equation}

The parameters $D_m^{12}$ and $\Delta\hat{V}_m$ can be
fixed from fits of the data to Eq.~(\ref{eq:correl}). The hybrid potentials
$\hat{V}_m$ can in principle also be determined
independently~\cite{huntley2}. We leave this for future high precision
studies on anisotropic lattices. For the time being, we
evaluate the integrals Eqs.~(\ref{ce_1})--(\ref{md_4}) numerically.
The interpolation method used for this purpose is inspired by the
multi-exponential result of the spectral decomposition,
Eq.~(\ref{eq:correl}).

\subsection{Integration errors}
The ${\mathcal O}(v^2)$ potentials are extracted from
integrals over correlation functions [see
Eqs.~(\ref{ce_1})--(\ref{md_4})] that depend
on the interaction time $t$ in a multi-exponential
way. In the
following, $C_i(t)$ will denote the
two-point function which has to be
integrated out in order to determine a potential $\hat{V}_i$
at a given
value of ${\mathbf R}$. For $i=1, 2$, $C_i(t)$ will be weighted by an
additional factor $t$ [Eqs.~(\ref{ef_1})--(\ref{ef_2})],
for $i=b,c,d,e$ by $t^2$ [Eqs.~(\ref{md_1})--(\ref{md_4})]. Two
different methods of interpolating $C_i(t)$ in between the discrete
$t$-values have been adopted:

1.\ We perform local exponential interpolations, which are
expected to yield the most reliable results:
\begin{equation}
C_i(t')=C_i(t)e^{-B_i(t)(t'-t)},\quad
B_i(t)=\ln\left[\frac{C_i(t)}{C_i(t+1)}\right]
\end{equation}
for $t\leq t'<t+1$ and $C_i(t)C_i(t+1)>0$. Due to the
multi-exponential character of the correlation function
(or statistical fluctuations) the sign might change within
the given interval. Thus, for
$C_i(t)C_i(t+1)\leq 0$, we interpolate
linearly,
\begin{equation}
C_i(t')=C_i(t)+[C_i(t+1)-C_i(t)](t'-t).
\end{equation}
For $C_1(t)$ and $C_2(t)$ quadratical interpolations are performed
within the interval $0\leq t'<\frac{1}{2}$
to account for $C_1(0)=C_2(0)=0$, where we
demand continuity of the interpolating function and its derivative
at $t=\frac{1}{2}$.

2.\ In order to estimate systematic integration errors,
simple linear interpolations of the data have been performed.

All statistical errors have been bootstrapped. For each potential
$\hat{V}_i$, numerical
integration has been performed up to a
value $t=\tau_i$, with $\tau_i$ chosen such that the result is
stable (within statistical accuracy) under the replacement $\tau_i
\rightarrow \tau_i-1$ for all ${\mathbf R}$.
Systematic cut-off errors have been estimated from the exponential
tail of fits to large-$t$ data points and have been found
to be negligible in all cases when
compared to the statistical
error from the numerical integration.
Typically, $\tau_i$ came out to be 6--8 lattice units~\cite{phd}.
For some of the correlation functions, the disconnected part
had to be subtracted. Its value has been estimated by averaging
data points within a range of $t$-values, $t\geq \tau^a_i$,
under variation of $\tau^a_i$
until a plateau was reached. Subsequently, the resulting value
has been subtracted from the correlator, before proceeding
to interpolation methods 1 and 2. In all cases, $\tau^a_i-1\leq
\tau_i\leq\tau^a_i$ was found, in support of self-consistency
of the method.

In what follows, we always state the result from the exponential
interpolation
method with the bootstrapped statistical error and
a systematic error that corresponds to the difference
between the results obtained from the two methods.
We find the systematic error to be the dominant source of
uncertainty, which can only be reduced by decreasing the temporal
lattice spacing.

\section{The matching problem}\label{sec3b}
\subsection{Renormalization of lattice operators}
The relativistic corrections to the static
potential are computed from amplitudes of
correlation functions
rather than from eigenvalues of the transfer
matrix. Therefore, they undergo renormalization.
This is in accord with the fact that
the electric and magnetic ears
explicitly depend on the
lattice scale $a$ and, thus, discretization.

As in the low energy regime of interest the renormalization constants
are likely to receive relevant corrections beyond one-loop
perturbation theory, we
apply the nonperturbative HM renormalization
prescription~\cite{huntley} [cf.\ Eq.~(\ref{defcor})].
The HM procedure is similar to the mean field
inspired tadpole improvement program,
advocated in Ref.~\cite{tadpole}. However, instead of
just dividing
correlators of eared Wilson loops
by the square of the average plaquette,
\begin{equation}
U_{\Box}=\frac{1}{6V}\left\langle\sum_{n,\mu>\nu}\frac{1}{3}
\mbox{Re\,Tr} P_{\mu,\nu}(n)\right\rangle,
\end{equation}
a more
sophisticated combination is chosen;
the various orientations of ears are taken into account,
such that the remaining renormalization constants
will only differ from identity on a three-loop
[$1 +{\mathcal O}(g^6)$] or two-loop
[$1+{\mathcal O}(g^4)$] level for operators involving
magnetic and electric ears, respectively. 
Details are discussed in Ref.~\cite{wachter}.

In the case of tadpole improvement, each electric or magnetic field $gE$ or
$gB$, appearing within the correlators of Eqs.~(\ref{ce_1})--(\ref{md_4}),
is multiplied by a constant $Z_{\mbox{\scriptsize tadpole}}=
1/U_{\Box}$. The difference between this procedure and the
HM scheme can be parameterized in terms of
a ratio,
\begin{equation}\label{qdef}
Q=\frac{\left\langle{\cal
P}\left[W(\Pi_i+\Pi_i^{\dagger})\right]\right\rangle}
{2\left\langle W\right\rangle U_{\Box}},
\end{equation}
that depends on the orientation of the ear, $i$, as well as
${\mathbf R}$ and $t$. In Fig.~\ref{renorm1} we display
this ratio for all independent color-electric and -magnetic components
at $R=4$ and $\beta=6.2$ as a function of $t$. No significant
$t$-dependence is observed, such that the renormalization
constants within integrals Eqs.~(\ref{ce_1})--(\ref{md_4})
factorize\footnote{This behavior is expected from the spectral decomposition
of the correlation function~\cite{wachter}, where the renormalization
factor corresponds to a constant, $1/g_{00}^i$, that only depends
on ${\mathbf R}$ and the specifications of the ear $i$. Any residual
time dependence has to
be attributed to the finiteness of $\Delta t_i$.};
all $Q$-factors saturate into
asymptotic values for $t\geq 4$, within
statistical errors of ${\mathcal O}(10^{-4})$.
In Fig.~\ref{renorm2}, the $R$ dependence of $Q$ is depicted
for on-axis separations. In the case of magnetic ears, the result appears
to be rather insensitive to the component and $R$.
For electric ears, $Q$ changes significantly with $R$ as well as the
component. However, at large separations, the electric components
approach a common value too. We call these plateau values $Q_B$ and
$Q_E$, respectively. In Table~\ref{rent}, we compare
$Z_{\mbox{\scriptsize tadpole}}$ with the
HM constants $Z_B=Z_{\mbox{\scriptsize tadpole}}/Q_B$
and $Z_E=Z_{\mbox{\scriptsize tadpole}}/Q_E$
for the two $\beta$-values. The magnetic renormalization constant
differs only by less than 1~\% from the tadpole value while
for the electric fields this difference amounts to 3--4~\%.
As expected, the disagreement decreases with the lattice spacing $a$.
We find it interesting to notice that the factors $Q$
are smaller than ratios of nonperturbative values~\cite{alpha}
of the coefficient of the clover term
within the Sheikholeslami-Wohlert fermion action~\cite{clover} and its
tadpole guesses. Also, the correction
goes opposite in the present case; $Z_B$ and $Z_E$ come out
to be smaller than the tadpole estimate $Z_{\mbox{\scriptsize
tadpole}}$.

Direct numerical checks of the accuracy of the HM
approach are possible
in two ways, namely (i) by varying the lattice resolution $a$ and
a scaling test of the results\footnote{Due to the running
of the matching constants to the full theory with the lattice scale,
residual scaling violations for the SD potentials
are expected from Eqs.~(\ref{v1sca}) and (\ref{v2sca}).}
and (ii) by comparing the data with
predictions from the exact Gromes and BBMP
relations, Eqs.~(\ref{grom})--(\ref{bram}),
between SD or MD potentials and the static
potential (which does not undergo renormalization),
\begin{eqnarray}
V_{2,\mbox{\scriptsize ren}}'(\pi/a;r)-
V_{1,\mbox{\scriptsize ren}}'(\pi/a;r)
&=&V_0'(r),\\
V_{b,\mbox{\scriptsize ren}}(r)
+2V_{d,\mbox{\scriptsize ren}}(r)
&=&\frac{r}{6}V_0'(r)-\frac{1}{2}V_0(r),\\
V_{c,\mbox{\scriptsize ren}}(r)
+2V_{e,\mbox{\scriptsize ren}}(r)
&=&-\frac{r}{2}V_0'(r).
\end{eqnarray}
In Fig.~\ref{v12} we check our data on $V_2'-V_1'$
against the force, obtained from fits
to the static potential,
$V_0(r)$ according to the parametrization Eq.~(\ref{eq_v0_3}) below.
As can be seen, the two data sets scale onto each
other and reproduce the static force.
The BBMP relations
are only satisfied on a qualitative level
as Figs.~\ref{vbd} and \ref{vce} demonstrate;
substantial lattice artefacts are responsible for deviations
from the expectations in the region of small $R$.

\subsection{Matching constants}
In order to calculate the matching constants between the effective
nonrelativistic Hamiltonian of
Eqs.~(\ref{ham})--(\ref{mdpo})
and QCD, we require values for the strong coupling constant
at scales $q=\pi/a$ and $q=m_b, m_c$ in a given
renormalization scheme. We decide to use the
``$V$'' scheme of Ref.~\cite{brodsky}, and compute the running coupling
from the average plaquette
as suggested in Ref.~\cite{tadpole},
\begin{equation}
\alpha_V^{-1}(q)= -4\pi\left[\frac{c_1}{\ln U_{\Box}}+2b_0
\ln\left(\frac{\pi}{aq}\right)+0.1058\right],
\end{equation}
where $c_1=1/3$ and $b_0=11/(16\pi^2)$ for SU(3) gauge
theory.

We use the plaquette values of Table~\ref{rent}
and Eqs.~(\ref{c_2})--(\ref{c_4}) to obtain the matching
constants $c_2(\mu,m)$ and $c_3(\mu,m)$ at $\beta=6.0$ and $\beta=6.2$,
listed in Table~\ref{ren2t}. We have assumed $m_b=4.7$~GeV
and $m_c=1.3$~GeV for the bottom and charm quark masses,
respectively. The value $\sqrt{\kappa}=468$~MeV has been used to fix
the lattice scale. These values
are obtained from the quarkonia
spectroscopy below.
Since $\alpha_V(q)$ depends only logarithmically
on $q$, accurate values for $a$, $m_c$ and $m_b$ are not required.

In the case of bottomonium
all constants turn out to be reasonably close to 1, such that
one-loop perturbation theory appears to be trustworthy.
However, for $m=m_c$, the size of the constants
indicates that higher order corrections cannot be neglected at
present lattice spacings $a\ll\pi/m_c$. Increasing the lattice spacing
would result in increased lattice artefacts as well as a larger
uncertainty in the renormalization factors $Z_B$ and $Z_E$ that relate
the lattice potentials to their continuum counterparts. In order to
achieve a reasonable balance between the uncertainties involved in
both matching procedures for the charmonium family,
improved lattice actions~\cite{imp} would have to
be considered.

\section{Results on the potentials}
\label{sec4}
We present numerical results and parametrizations on the static
potential, the relativistic corrections to the central potential,
and the SD and MD potentials. We compare the short range
SD potentials $V_2'$, $V_3$ and $V_4$ and the MD
potential $\hat{V}_c$ to lattice perturbation theory.
The short range SD potentials might provide another access to the
determination of the QCD coupling $\alpha_V(\pi/a)$,
quite in the spirit of the role of the fine structure
constant in the analysis of atomic level splittings.

\subsection{The static potential}
The lattice potential
$\hat{V}_0({\mathbf R})$ has been computed from smeared Wilson loops
by use of the method described in Ref.~\cite{schlicht}.
Our general strategy is to derive interpolating
parametrizations of the
lattice data points which will enable us to compare
the results to continuum expectations.
Weak coupling continuum and lattice predictions
on the potentials have
been presented in Sec.~\ref{sec2}
[Eqs.~(\ref{vp0})--(\ref{vpc}) and Eqs.~(\ref{aa6})--(\ref{abe6}),
respectively],
such that we can correct the
lattice data for the differences between
both tree-level forms before attempting to fit them to a
continuous parametrization.
Let
\begin{equation}
\hat{V}_{0,\mbox{\scriptsize cont}}(R)=\hat{V}_0({\mathbf R})-
g\delta \hat{V}_0({\mathbf R})-\hat{C}_V,
\label{eq_v0_2}
\end{equation}
with
\begin{equation}
\delta \hat{V}_0({\mathbf R}) =
-4\pi G_L({\mathbf R})+\frac{1}{R}
\label{eq_v0_1}
\end{equation}
be the tree-level corrected static potential.
$G_L({\mathbf R})$ is the lattice gluon propagator
of Eq.~(\ref{a10}).
The static lattice potential is fitted
to the five-parameter ansatz [including $g$ and $\hat{C}_V$
of Eq.~(\ref{eq_v0_2})
as fit parameters], 
\begin{equation}
\hat{V}_{0,\mbox{\scriptsize cont}}(R)
= K R -\frac{e}{R}+\frac{\hat{f}}{R^2}
\label{eq_v0_3}
\end{equation}
with string tension $K=\kappa a^2$
and Coulomb coefficient $e$. The $1/R^2$
correction, that accounts
for the running of the coupling, is not meant
to be physical but has been introduced
to effectively parameterize
the data within the given range of $r$-values.
The resulting parameter values are displayed
in Table~\ref{et2}\footnote{The reduced $\chi^2$-values
stated in the table do not take account of correlations between
data points obtained at different $R$.}.
For technical reasons related to the link integration procedure,
only potential values for $R\ge \sqrt{2}$ have been obtained,
such that the fits do not include $R=1$.

In Fig.~\ref{v0}, the
potential $V_{0,\mbox{\scriptsize cont}}$ from both
$\beta$-values is
displayed in physical units (as obtained from
$\sqrt{\kappa}=468$~MeV),
together with a fit curve that corresponds to the (averaged) 
values of fit parameters $e=0.321(6)$ and
$f=a\hat{f}=0.0082(8)/\sqrt{\kappa}$.
As can be seen, the two data
sets scale nicely onto each other. Violations of
rotational invariance
are removed by the correction method, even at very small values of
$R$, and the data is well described by the parametrization over the
entire $r$-range.

\subsection{Corrections to the central potential}
Fits of $\nabla^2\hat{V}_a^E$ to the parametrization of
Eq.~(\ref{vapa}),
\begin{equation}\label{eqn1}
\nabla^2\hat{V}_a^E=\hat{C}_a^E-\frac{\hat{b}}{R},
\end{equation}
with parameters $\hat{C}_a^E$ and $\hat{b}$ have been performed.
The resulting potential $\nabla^2V_a^E-C_a^E$ is shown
in Fig.~\ref{vae}.
Results on $b$ are displayed in the first row of Table~\ref{et3}.
Systematic errors from the integration procedure are included
in square brackets. We find the values
$\chi^2/N_{DF}=0.5$ and $\chi^2/N_{DF}=2.0$ at
$\beta=6.0$ and $\beta=6.2$, respectively, for the fit range
$R\geq\sqrt{2}$. These
$\chi^2$-values refer only to the statistical errors.
The fitted curve that
corresponds to the averaged value $b=(0.86\pm 0.05$~GeV$)^2$ is
included into the figure.
{}From Eq.~(\ref{vaesca}) and the matching constants of
Table~\ref{ren2t}, we expect scaling violations of about 10~\% between
the two data sets. Apart from the region $r<0.15$~fm, which is
polluted by lattice
artefacts, this effect cannot be resolved within
statistical accuracy.

In Fig.~\ref{vab}, we display $\nabla^2\hat{V}_a^B$ in lattice units
at $\beta=6.2$, where the error bars of this plot refer to the statistical
uncertainty only. The $\beta=6.0$ data exhibits
the same qualitative behavior.
The large-$R$ data can be parameterized by a constant. Deviations
from this constant at small $R$-values, which are hidden within the
systematic uncertainty of the integration, can be due either to lattice
artefacts or to a tiny $\delta$-like admixture that one might expect
from Eq.~(\ref{vabsca}).
The numerical values (with statistical and systematic errors) are
$\hat{C}_a^B=-1.02(1)(27)$ and  
$\hat{C}_a^B=-0.93(1)(24)$ at $\beta=6.0$ and $\beta=6.2$,
respectively.
These values have to be related to
$\hat{C}_a^E=-1.00(2)(8)$ and  
$\hat{C}_a^E=-0.92(1)(8)$, such that
$C_a^B = C_a^E$ within errors.

We conclude that the corrections to the central potential agree
reasonably well with the expectations of Eq.~(\ref{vapa}), with a
parameter $b\approx 4\kappa\approx 0.9^2$~GeV$^2$.
The strength of the effective Coulomb coupling is increased by about 2~\%
in case of the $\Upsilon$ family and 35--40~\% for $J/\psi$ states,
due to these correction terms. The self-energy type
contributions to $\nabla^2 V_a^E$ and $\nabla^2 V_a^B$
cancel each other at the present level of statistical accuracy.

\subsection{Spin-dependent potentials}
Our results on the first spin-orbit potential $V_1'$ are displayed in
Fig.~\ref{v1}. The two data sets show
approximate scaling behavior. In addition to a constant long range
contribution, $-K$, we find an
attractive short range contribution that 
can be fitted to the Coulomb-like ansatz of Eq.~(\ref{v1pa}),
\begin{equation}\label{eqn2}
\tilde{V}_1'(r)=-\frac{h}{R^2}-K,
\end{equation}
in agreement with our SU(2)
investigation~\cite{wachter}.
For these one-parameter
fits we have constrained the constant long range part to the value of
the string tension, as obtained from the static potential. We find
the values $h=0.071(12)$ and $h=0.065(9)$ for $\beta=6.0$ and
$\beta=6.2$, respectively.
As expected from Eq.~(\ref{v1sca}), $h$ tends to decrease with $\beta$.

Taking the Gromes relation and the running coupling improved effective
parametrization
of Eq.~(\ref{eq_v0_3}) into account, we expect,
\begin{equation}
\label{v2exp}
V_2'(r)=\frac{e-h}{r^2}-\frac{2f}{r^3}.
\end{equation}
Note, that we have added a $1/r^3$-term to the expectation
Eq.~(\ref{v1pa}) which accounts for a weakening of the effective
coupling with decreasing source separation.
{}From Eq.~(\ref{exv31}) we expect the parametrization,
\begin{equation}
\label{v3exp}
V_3(r)=\frac{3(e-h)}{r^3}-\frac{8f}{r^4},
\end{equation}
to approximate $V_3$.

Prior to comparing the data to the above continuum parametrizations, we
attempt to correct for lattice artefacts. For this purpose we define,
\begin{eqnarray}
\delta\tilde{V}_2'({\mathbf R})&=&
\frac{4\pi}{C_Fg^2}\tilde{V}'_{2,\mbox{\scriptsize tree}}({\mathbf R})-
\frac{1}{R^2},\\
\delta\tilde{V}_3({\mathbf R})&=&
\frac{4\pi}{C_Fg^2}\tilde{V}_{3,\mbox{\scriptsize tree}}({\mathbf R})-
\frac{3}{R^3}.
\end{eqnarray}
The lattice tree-level potentials
$\tilde{V}'_{2,\mbox{\scriptsize tree}}$ and
$\tilde{V}_{3,\mbox{\scriptsize tree}}$ are defined in
Eqs.~(\ref{v2p})--(\ref{latv32}). We then correct
for lattice artefacts,
\begin{eqnarray}\label{eqn3}
\tilde{V}'_{2,\mbox{\scriptsize cont}}(R)&=&\tilde{V}_2'({\mathbf R})
-g_2\delta\tilde{V}_2'({\mathbf R}),\\
\tilde{V}_{3,\mbox{\scriptsize cont}}(R)&=&\tilde{V}_3({\mathbf R})
-g_3\delta\tilde{V}_3({\mathbf R}),
\end{eqnarray}
and fit the potentials to the following ans\"atze:
\begin{eqnarray}
\tilde{V}'_{2,\mbox{\scriptsize cont}}(R)&=&\frac{c_2}{R^2}
-\frac{2\hat{f}_2}{R^3},\\
\tilde{V}_{3,\mbox{\scriptsize cont}}(R)&=&\frac{3c_3}{R^3}
-\frac{8\hat{f}_3}{R^4},\label{eqn4}
\end{eqnarray}
where $g_i$, $c_i$ and $\hat{f}_i$ are fit parameters. The resulting
parameter values are shown in Table~\ref{et4}. Again, the
$\chi^2$-values refer to the statistical errors only.

The fitted values $\hat{f}_2$ and $\hat{f}_3$ are in agreement with
$\hat{f}$ as extracted from the static potential. Also, $c_2$ and $c_3$
agree with $e-h$ as computed from $\hat{V}_0$ and $\tilde{V}_1$
reasonably well. Only the coefficients of the correction terms, $g_2$
and $g_3$, come out to be about a factor 2 smaller than in the case of
the static potential. The spin-orbit potential
$V'_{2,\mbox{\scriptsize cont}}$
and the spin-spin potential
$V_{3,\mbox{\scriptsize cont}}$ are displayed in
Figs.~\ref{v2} and \ref{v3}, respectively,
together with
the theoretical
expectations.
In both cases, we observe reasonable agreement between data and
expectation and
the two data sets from the
different $\beta$-values scale nicely onto each other,
after we have corrected for tree-level lattice artefacts.

In Fig.~\ref{v4}, the spin-spin potential $\tilde{V}_4$ is displayed in
lattice units for the two $\beta$-values.
An oscillatory behavior is observed which is similar to that of the
lattice $\delta$-function, expected at tree-level, Eq.~(\ref{latv4}).
Moreover, the two data sets nearly coincide with each other,
in distinct violation of scaling.
Corrections to the $\delta$-function, which might scale
with an appropriate dimension, should account for
the differences between the two data sets at small $R$.

\subsection{Momentum-dependent potentials}
We intend to compare the MD potentials to
Eqs.~(\ref{vbpa})--(\ref{mdpa}). Since, in accord
with these expectations, the MD potentials
are rather small, compared to the SD potentials, the data suffers more
from statistical noise, and
we do not attempt to perform fully independent fits.
In addition, we neglect running coupling effects that have been
parameterized by $f$ in the case of $V_0$, $V_2'$ and
$V_3$. We have to subtract the self-energy related
constants $\hat{C}_b$ and
$\hat{C}_d$
from the data points on $\hat{V}_b$ and $\hat{V}_d$, respectively,
prior to scaling the data sets onto each other. We also
correct $\hat{V}_b$ and $\hat{V}_c$ for tree-level lattice
artefacts. For this purpose we define,
\begin{eqnarray}
\delta\hat{V}_b({\mathbf R})&=&
\frac{4\pi}{C_Fg^2}\hat{V}_{b,\mbox{\scriptsize tree}}({\mathbf R})-
\frac{2}{3R},\\
\delta\hat{V}_c({\mathbf R})&=&
\frac{4\pi}{C_Fg^2}\hat{V}_{c,\mbox{\scriptsize tree}}({\mathbf R})+
\frac{1}{2R}.
\end{eqnarray}
The tree-level lattice expectations for $\hat{V}_b$ and $\hat{V}_c$ 
can be computed from Eqs.~(\ref{latvb}) and
(\ref{latvc}).

We fit the data to the following parametrizations,
\begin{eqnarray}\label{eqn5}
\hat{V}_b({\mathbf R})&=&\frac{2e}{3R}-\frac{1}{9}KR+\hat{C}_b+
g_b\delta\hat{V}_b({\mathbf R}),\\
\hat{V}_c({\mathbf R})&=&-\frac{e}{2R}-\frac{1}{6}KR+
g_c\delta\hat{V}_c({\mathbf R}),\\
\hat{V}_d({\mathbf R})&=&-\frac{1}{9}KR+\hat{C}_d,\label{eqn6}
\end{eqnarray}
where we have constrained the parameters $e$ and $K$ to the values,
obtained from the fit to the static potential.

The resulting parameter values are listed in Table~\ref{et5}.
In accord with the BBMP relation Eq.~(\ref{bram2}), we find
$-2\hat{C}_b-4\hat{C}_d=\hat{C}_V$ (within errors) for both $\beta$-values.
The corrected potentials,
\begin{eqnarray}
\hat{V}_{b,\mbox{\scriptsize cont}}(R)&=&
\hat{V}_b({\mathbf R})-g_b\delta\hat{V}_b({\mathbf R})-\hat{C}_b,\\
\hat{V}_{c,\mbox{\scriptsize cont}}(R)&=&
\hat{V}_c({\mathbf R})-g_c\delta\hat{V}_c({\mathbf R}),
\end{eqnarray}
as well as $\hat{V}_d-\hat{C}_d$ and
$\hat{V}_e$ are displayed in
Figs.~\ref{vb}--\ref{ve}. 
The expectations
Eqs.~(\ref{vbpa})--({\ref{mdpa}) are included as well (solid curves).
The $\chi^2/N_{DF}$ values of the above fits are larger
than 1 for $\hat{V}_b$ and $\hat{V}_c$, which means
that the correction for lattice artefacts of these potentials is
not as successful as it has been
in the case of $\tilde{V}_2'$ and $\tilde{V}_3$.
This can be understood from the fact that the MD potentials
are more strongly affected by the infrared behavior of the gluon
propagator, such that higher order corrections might be important.
$\hat{V}_e$ shows substantial lattice artefacts too (Fig.~\ref{ve}).
In the case of $V_b$ the small-$R$ data lies below the curve,
indicating that the $1/r$
coefficient $2e/3$ might have been overestimated. This effect cannot
be understood in terms of the
tiny $f/r^2$ correction that has been omitted.
However, by allowing for a $1/r$-term with a coefficient of
about $e/8$ in $V_d$, the expectation can be brought into agreement with
the data. Within $-V_c$ the $1/r$ coefficient $e/2$ appears to be
slightly underestimated.

We conclude that the data is in qualitative
agreement with the expectations Eqs.~(\ref{vbpa})--(\ref{mdpa}),
although a quantitative comparison
fails as there are indirect indications that $V_d$ and $V_e$ 
might contain
small Coulomb-like contributions, in addition to the linear
term.

\subsection{Comparison with perturbation theory}
\label{FAE}
In Figs.~\ref{latt_v2}--\ref{latt_vc}, we focus on the small-$R$
behavior of the SD
potentials $\tilde{V}_2',\ldots, \tilde{V}_4$ and the MD potential
$\hat{V}_c$. We show only the $\beta=6.2$ results, which are in
qualitative agreement
with those obtained at $\beta=6.0$. Besides the data points, 
the figures include
the tree-level perturbative expressions of
Eqs.~(\ref{v2p})--(\ref{latv4}) and
Eq.~(\ref{latvc}). The normalization constants
$c=C_F\alpha_s$ have been
obtained from fits to the first seven data points. 
$\tilde{V}'_2$ and $\tilde{V}_3$ are
well described by these one-parameter fits and
deviations of the data from a continuous curve can be
understood in terms of this lattice expectation. For $\tilde{V}_4$ as
well as $\hat{V}_c$, agreement is only achieved on a qualitative level.
The fit parameters are displayed in Table~\ref{et6}. 

{}From the analysis
of the static potential, we expect $c=e-h\approx 0.25$, compared to
the tree-level lattice expectations
$c=0.106$ and $c=0.102$ for $\beta=6.0$ and $\beta=6.2$,
respectively, determined from the lattice coupling
$\alpha_s=3/(2\pi\beta)$.
In agreement with the
perturbative expectation, all fitted $c_i$ decrease with
increasing
$\beta$. We find $c_c$ to be
about 5 times as large as the
naive tree-level value; this factor reduces to 2.4 in the case of
$\tilde{V}_2'$ and 1.9 for $\tilde{V}_3$ and $\tilde{V}_4$ as the
relevant gluon momenta within these potentials are larger and
thus more perturbative.

In order to investigate if remaining differences between data
points and renormalized tree-level expectations can be explained in
terms of higher order
perturbative corrections, we attempt to model running coupling
effect.
The only additional diagrams that
contribute to $V_0$ at ${\mathcal O}(g^4)$ on the lattice (and
in the continuum) are one-loop corrections to the gluon self-energy.
The renormalization of the coupling, emanating from these diagrams, has
been computed on the lattice for on-axis separations of the
sources~\cite{paffuti,karsch}. One can account for this correction by
building in a running coupling constant $\alpha({\mathbf q})$ into the
gluon propagator of Eq.~(\ref{a10}),
in momentum space.
Instead of attempting to compute the correct lattice sum, we model
this effect by the corresponding continuum expression,
\begin{equation}\label{running}
\alpha(t)=\frac{1}{4\pi b_0t}\left[1+\frac{b}{t}\ln t\left(
1+\frac{b}{t}\right)\right]^{-1},
\end{equation}
with
$t=\ln\left(\hat{q}^2/\Lambda^2\right)$,
$b=b_1/b_0^2$, $b_0=11/(16\pi^2)$, $b_1=102/(16\pi^2)^2$,
where we replace $q^2$ by its lattice counterpart
$\hat{q}^2=4\sum_i\sin^2(q_i/2)$~\cite{huntley,klassen}.
$\Lambda$ is a QCD scale
parameter that can
be related to the usual schemes via perturbation
theory~\cite{Pantaleone,fischler,billoire}.
The difference between the correct one-loop lattice expression
of Ref.~\cite{karsch} and Eq.~(\ref{running}) with $b=0$ is small. 

In the continuum, contributions that
appear in addition to a pure renormalization of the gluon
propagator can be
resummed into a single running coupling, using renormalization group
arguments. On the lattice
rotational invariance is broken
and the direction of ${\mathbf q}$ enters in
addition to its absolute value;
hence such arguments do not apply.
Bearing this in mind, we will nonetheless attempt to model
higher order perturbative effects by the continuum
running coupling of Eq.~(\ref{running}).

In the case of the SD potentials $V_2',\ldots, V_4$ not only the gluon
self-energy contributes to ${\mathcal O}(g^4)$ but also exchange
diagrams between the ears, incorporating a three gluon
vertex. In the continuum, these can be resummed into an effective
running coupling. Due to this resummation, the scale parameters
$\Lambda_i$ (for $V_i^{(\prime)}$) can
differ from each other.
However, one finds
$\Lambda_2=\Lambda_V$~\cite{Pantaleone,fischler,brodsky}.

To remove the
unphysical pole at $q=\Lambda$, an infrared
protection can be built into the propagator by substituting $t$ by
$t_d=\ln(q^2/\Lambda^2+d^2)$ with a constant $d$.
The smallest momentum on the lattice is $q=\pi/(aL_{\sigma})$. We
choose $d^2=\max(0,e-\pi^2/(aL_{\sigma}\Lambda)^2)$, where $e$ is the Euler
constant, to guarantee
$t\geq 1$; $d^2$ is negligible
at large momenta $q\approx 1/a$. Notice, that
within the SD potentials
the infrared region is suppressed by
powers of $q$, such that the results are rather
robust with respect to the
choice of $d$ or other specific details of the protection scheme.

Fits of the one-
and two-loop running coupling 
improved expressions
to the first 4--8 data points of each potential have been
performed.
$\Lambda$ is the only free parameter within these fits.
The results of the one-loop fits to 7 data points
are included into Figs.~\ref{latt_v2}--\ref{latt_v4} (full circles).
The $\Lambda$-parameters remain stable against the variation of the
fit range within
errors. Since the data is described by the tree-level
formulae equally well, we are unable to decide at present whether the
deviations between expectation and data for $\hat{V}_4$
can be explained entirely
in terms of such higher order perturbative corrections.

In Tables~\ref{et7} and \ref{et8},
results on one- and two-loop estimates of $\Lambda$-parameters
are presented.
We observe scaling
between the two sets of $\Lambda$-parameters obtained at $\beta=6.0$ and
$\beta=6.2$. The two-loop values are about twice as
large as the corresponding one-loop values.
However, the (one- and two-loop) $\alpha_i(q)$-values at
a scale $q=\pi/a$ are consistent with each other.
We conclude that the one-loop $\Lambda$-values should
be considered as effective and not physically meaningful.

{}From our one-loop fits to $V_2'$, we obtain
$\alpha_V(\pi/a_{6.0})=0.131^{+7}_{-4}$ and
$\alpha_V(\pi/a_{6.2})=0.124^{+5}_{-4}$ at $\beta=6.0$ and
$\beta=6.2$, respectively. The corresponding two-loop results,
$\alpha_V(\pi/a_{6.0})=0.128^{+6}_{-8}$ and
$\alpha_V(\pi/a_{6.2})=0.121^{+5}_{-5}$,
are in nice agreement with these numbers, while from the
average plaquette~\cite{tadpole} we obtain
$\alpha_V(\pi/a_{6.0})=0.149$ and
$\alpha_V(\pi/a_{6.2})=0.138$. We conclude that
higher order perturbative corrections as well as
${\mathcal O}(a^2)$ discretization
errors, under which localized quantities like the plaquette are more
likely to suffer,
are responsible for the $2\sigma$--$3\sigma$
deviations between the $\alpha_V$-values,
determined from two different observables.

\section{Application to quarkonia spectra}
\label{sec5}
With the potentials
derived from quenched QCD we would now like to
predict experimental quarkonia levels. The spectrum
will be computed
numerically from the two-body Hamiltonian, the structure
of which will be summarized in the next subsection.

\subsection{The Hamiltonian}
Within the spectroscopy study,
we restrict ourselves to the equal mass case $m=m_1=m_2$.
The starting point is the Hamiltonian,
\begin{equation}\label{ham22}
H=2(m-\delta m)+H_0+\delta H_{\mbox{\scriptsize kin}}+
\delta H_{\mbox{\scriptsize cc}}+
\delta H_{\mbox{\scriptsize sd}}+\delta H_{\mbox{\scriptsize md}},
\end{equation}
where
\begin{equation}\label{ham23}
H_0=\frac{{\mathbf p}^2}{m}+\left[\overline{V}(r)-
\delta H_{\mbox{\scriptsize cc}}(r)\right],
\end{equation}
contains the Coulomb-like part within the relativistic correction to the
central potential.
We numerically solve the radial Schr\"odinger equation for $H_0$,
and treat
\begin{equation}\label{ham24}
\delta H_{\mbox{\scriptsize kin}}=-\frac{p^4}{4m^3},\quad
\delta H_{\mbox{\scriptsize cc}}=\frac{c_3(m)}{m}\pi e\delta^3(r),\quad
\delta H_{\mbox{\scriptsize sd}}=V_{\mbox{\scriptsize sd}},\quad
\delta H_{\mbox{\scriptsize md}}=V_{\mbox{\scriptsize md}}
\end{equation}
as perturbations~\cite{phd}.

For the particular parametrizations
Eqs.~(\ref{v0pa}) and (\ref{vapa}),
one obtains the central potential [Eq.~(\ref{cepo})],
\begin{equation}\label{ham25}
\overline{V}(r)=\kappa r-\frac{e}{r}+\frac{1}{8}
\left(\frac{c_3(m_1)}{m_1}+\frac{c_3(m_2)}{m_2}\right)
\left[\left(2\kappa-b\right)\frac{1}{r}+4\pi e\delta^3(r)\right].
\end{equation}
The perturbation $\delta H_{\mbox{\scriptsize cc}}$ is due to
the last term within this equation.
We have omitted the constants $C_V$, $C_a^B$ and $C_a^E$
from the above formula. The latter two of these contributions
cancel each other within the
statistical accuracy of our lattice
results while, as we shall see below,
$C_V$ can be absorbed into a redefinition of the quark masses.
The $\kappa/r$- and $\delta^3(r)$-terms have their origin
in the Darwin interaction while the $b/r$-term is due to
$\nabla^2V_a^E$. The mass dependent correction terms
explain the phenomenological flavor dependence of the
central quarkonium potential, as obtained from fits
to the spin-averaged charmonia and bottomonia spectra~\cite{qui,fulcher}.

Two-particle bound states can be classified by
a radial excitation $n$,
the orbital angular momentum $L$, the total spin $S=0,1$
(${\mathbf S}={\mathbf S_1}+{\mathbf S_2}$), and a
total angular momentum $J=L-S,L,L+S$
(${\mathbf J}={\mathbf L}+{\mathbf S}$).
Conventionally, the states are labeled by $n^{2S+1}L_J$ where
the letters $S,P,D,F$ are used for $L=0,1,2,3$, respectively.
From parametrizations Eqs.~(\ref{v0pa}), (\ref{v1pa}) and
(\ref{v3pa}), we find (for equal masses),
\begin{eqnarray}
V_{\mbox{\scriptsize sd}}(r)&=&\frac{1}{m^2}\left[
\left(-\frac{\kappa}{r}+ \frac{4c_2(m) (e-h) -e}{2r^3}\right)
{\mathbf LS}\right.\nonumber\\\label{ham26}
&+&3c_2^2(m)\frac{(e-h)}{r^3}T\\\nonumber
&+&\left.\frac{1}{4}\left(7c_2^2(m)-3\right)8\pi(e-h)\delta^3(r)\frac{\mathbf
S_1S_2}{3}\right],
\end{eqnarray}
with
\begin{eqnarray}
\frac{\mathbf S_1S_2}{3}&=&
\frac{1}{6}\left(S(S+1)-\frac{3}{2}\right),\\
{\mathbf LS}&=&\frac{1}{2}\left(J(J+1)-L(L+1)-S(S+1)\right),\\
T=R_{ij}S_1^iS_2^j&=&
-\frac{6({\mathbf LS})^2+3{\mathbf LS}-2S(S+1)L(L+1)}
{6(2L-1)(2L+3)}.
\end{eqnarray}
The one-loop
values of the coefficients $c_2(m)$ and $c_3(m)$ for $m=m_b$ and
$m=m_c$ at our lattice spacings can be found in
Table~\ref{ren2t}. The values of the parameters
$e$, $h$ and $b$ are listed
in Tables~\ref{et2} and \ref{et3}.

Based on the parametrizations
Eqs.~(\ref{vbpa})--(\ref{mdpa}), we find the MD correction
[Eq.~(\ref{mdpo})],
\begin{equation}\label{ham27}
V_{\mbox{\scriptsize md}}(r)=-\frac{\mathcal K}{6r}
-\frac{\mathcal E}{r}p^2-\left(\frac{\mathcal K}{6}
-\frac{\mathcal E}{2r^2}
\right)\frac{1}{r}L^2-\frac{\mathcal E}{r^3}i{\mathbf rp},
\label{mdsp}
\end{equation}
where
\begin{equation}\label{ham28}
{\mathcal K}
=\kappa\left(\frac{1}{m_1^2}+\frac{1}{m_2^2}-\frac{1}{m_1m_2}
\right),\quad{\mathcal E}=\frac{e}{m_1m_2},
\end{equation}
i.e.\ ${\mathcal K}$ is a dimensionless parameter
while ${\mathcal E}$ carries the
dimension $m^{-2}$. 
Note that a string of constant longitudinal electric field
with energy density $\kappa$~\cite{buchmuller},
connecting two pointlike particles with masses $m_1$ and
$m_2$, gives rise to the classical correction term
$-\frac{\cal K}{6}\frac{1}{r}L^2$, which appears in the above
$V_{\mbox{\scriptsize md}}$.
One obtains the following expectation value within wave functions
that obey the Schr\"odinger equation,
\begin{equation}
-\left\langle\frac{1}{r^3}i{\mathbf rp}\right\rangle
=2\pi\left\langle\delta^3(r)\right\rangle,
\end{equation}
such that Eq.~(\ref{mdsp}) 
can readily be treated as a perturbation.

We have neglected the
constants $C_d$ and $C_b$ of $V_d$ and $V_b$
from Eq.~(\ref{mdsp}). Inclusion of these
terms would result in a
correction,
\begin{equation}
\Delta V_{\mbox{\scriptsize md}}=-\frac{1}{4}\left[
\left(\frac{1}{m_1^2}+\frac{1}{m_2^2}\right)C_V
+2\left(\frac{1}{m_1}+\frac{1}{m_2}\right)^2C_b\right]p^2,
\end{equation}
where we have exploited the
relation $C_V=-2C_b-4C_d$
from the BBMP constraint, Eq.~(\ref{bram}).
Under the assumption that $|C_b|\ll 2|C_d|$
(which
is supported by our numerical results, the MAL picture and
tree-level perturbation theory), the above shift in the MD potential
can be absorbed into a
redefinition
of the quark masses of Eq.~(\ref{ham}) at the given order
of the Hamiltonian:
$m_i\rightarrow m_i+C_V/2$. 
Notice, that the parameter $m$
of Eq.~(\ref{ham22}) differs from that of
Eq.~(\ref{ham}) by this constant.
If we interpret the fit parameter $\hat{C}_V$, obtained at a
lattice spacing $a$, as the self-energy of a static quark,
$E^{\infty}_0(\mu=\pi/a)\approx a^{-1}\hat{C}_V/2$,
the combination $m-\delta m$ within Eq.~(\ref{ham22}) should approach
the heavy quark pole mass.

\subsection{Spectroscopy results}
The physical scale, the quark mass $m$
and the energy shift $\delta m$ have to be fixed from
experiment, before predictions can be made.
Note that the parameter $m$ within the Hamiltonian
Eqs.~(\ref{ham22})--(\ref{ham28}) is not the bare quark mass,
but contains the static quark self-energy.
The value of the dimensionful parameter $\kappa$
determines the scale.
We attempt to estimate the (presumably small)
impact of the
parameter $\delta m$ on our results.
Therefore, we follow two strategies:
we minimize the squared differences between our predictions
and experimental levels under the assumption $\delta
m=0$, with respect to $m$ and $\kappa$ (method 1). Alternatively, we
determine $m$ and $\kappa$ from minimizing
deviations from the splittings $M(n^{2S+1}L_J)-M(1^3S_1)$.
Subsequently, $\delta m$ is tuned to reproduce
the $1^3S_1$ experimental
state (method 2). The latter method results in
the ratios $\delta m_b/m_{b,2}\approx 0.04$ and $\delta m_c/m_{c,2}\approx
0.22$. However, for ratios of the scales,
determined by use of the two methods, we obtain
$(\kappa_2/\kappa_1)^{1/2}\approx 1.0025$ and
$(\kappa_2/\kappa_1)^{1/2}\approx 1.035$ for bottomonia and charmonia
states, respectively. The tiny size of the deviations of these ratios
from unity
can be understood from
the fact that a simple rescaling of the mass works rather well as the following
ratios indicate:
$(m_2-\delta m)/m_1\approx 1.0015$ and 1.012,
again for bottomonia and charmonia, respectively.
This illustrates that although $\delta m$
can be quite substantial, and thus the uncertainty in the quark mass
can be large, the impact of this parameter on the predicted
spectrum is negligible. Hence, we adopt method 1,
i.e.\ we set $\delta
m=0$ and allow for two free parameters, $m$ and $\kappa$.

In the case of bottomonia, ten states have been observed and we
minimize our data with respect to all these states. The $B\overline{B}$
threshold is at about 10.55 GeV.
For charmonia, we chose to optimize the spectrum
only with respect to the seven states below 
the $D\overline{D}$ threshold at about 3.7~GeV.
The results for $\beta=6.0$ and $\beta=6.2$ are displayed in
Tables~\ref{spect1t} and \ref{spect2t} for bottomonia and
charmonia, respectively\footnote{
The effect of the statistical errors of the fit parameters on the
spectrum is negligible, in comparison to the systematic uncertainties
of the approximation, particularly those of the matching constants
$c_i(\mu,m)$. For this reason, we do not attempt to include
any errors into the tables.}.
We find
agreement on the level of about 2--3 MeV between the results obtained
at these two lattice spacings for bottomonia, compared
to about 10 MeV for charmonia. The differences are likely
to reflect the uncertainties in the matching coefficients $c_2(m)$ and
$c_3(m)$ which increase with decreasing quark mass.

The $\beta=6.2$ results are compared to experiment in
Figs.~\ref{bottom} and \ref{charm}. In order to estimate
the effect of quenching, we have included the results
obtained for the parameter value $e=0.40$, which is our estimate,
based on Ref.~\cite{SESAM}, of the
value one might obtain with three active
flavors of sea quarks.
In the case of bottomonia
states, an average deviation between prediction and experiment of
12.4~MeV at $\beta=6.0$ and 12.8~MeV at $\beta=6.2$ is observed. With
a value $e=0.40$, this deviation is reduced to 9.5~MeV. For charmonia, we
obtain an average deviation of 22.0~MeV for both $\beta$-values. The
parameter choice $e=0.40$ changes this to 23.0~MeV, indicating that the
charmonium spectrum is rather insensitive towards quenching effects on
the running of the QCD coupling.
This can be understood from the fact that the wave functions are
broader, such that the spectrum is less affected by short distance
physics.

From our fit to the bottomonium spectrum, we obtain the following
parameter values, both at $\beta=6.0$ and $\beta=6.2$,
\begin{equation}
\sqrt{\kappa}=468 \mbox{~MeV},\quad m_b=4.68 \mbox{~GeV}.
\end{equation}
The above string tension value yields a Sommer
scale~\cite{sommer} $r_0\approx
0.49$~fm. $r_0$ denotes the distance at which the
condition $r^2dV_0/dr=1.65$ is satisfied.
With $e=0.40$, we find $\sqrt{\kappa}=452$~MeV and
$m_b=4.72$~GeV.
The scale $r_0$ remains unaffected
under this change in $e$.
From the charmonium spectrum, we find,
\begin{equation}
\sqrt{\kappa}=450(4) \mbox{~MeV},\quad m_c=1.33(1) \mbox{~GeV}.
\end{equation}
The errors correspond to the variation between the results obtained at
the two $\beta$-values. In comparison, the sea quark model
with $e=0.40$
yields $\sqrt{\kappa}\approx 440$~MeV and $m_c\approx 1.38$~GeV.
The above results are consistent with pole masses
$m^{\mbox{\scriptsize pole}}_b=4.7(2)$ and $m_c^{\mbox{\scriptsize
pole}}=1.4(2)$~GeV, where the
errors are estimated from the $\delta m/m$-ratios.

From the fit to ten bottomonium states, we find lattice spacings
$a^{-1}\approx 2.1$~GeV and $a^{-1}\approx 2.9$~GeV for the two $\beta$-values,
respectively, which are in reasonable agreement with estimates from the
light hadron spectrum. If we forced the average of the $2^3S-1^3S$ and the
$\overline{1^3P}-1^3S$ splittings to coincide with the experimental 
counterpart,
as it is normally done in NRQCD studies, we obtain $a^{-1}\approx 2.5$~GeV
and $a^{-1}\approx 3.4$~GeV, respectively, which is in agreement with NRQCD
estimates~\cite{nrqcd}. As a result, however, the $2^3P$ masses would
come out to be significantly heavier than in experiment.

In order to investigate the reliability of the nonrelativistic
approximation, we have computed average radii and velocities of
various quarkonia states. The results are displayed
in Table~\ref{veloc}. The heavy quark velocities within bottomonia
range from $\langle v^2_b\rangle=0.07$ to $\langle v^2_b\rangle
= 0.11$, while for charmonia we obtain the interval, $0.27<\langle
v^2_c\rangle<0.52$. The radial bottomonia wave functions
$g_{nl}(r)$ [$\psi_{nlm}(r,\Omega)=[g_{nl}(r)/r]\,Y_{lm}(\Omega)$] are
shown in Fig.~\ref{wave}. 

We attempt to estimate the approximate size
of ${\mathcal O}(v^4)$ corrections, using the ratio
$R=\langle v^4\rangle/\langle v^2\rangle$ and find, $R_b\approx 0.1$
and $R_c\approx 0.4$. Under the assumption that the coefficients of
such corrections have the same size as those of the ${\mathcal
O}(v^2)$ corrections, we estimate an uncertainty of 4 and 15~MeV
for bottomonia and charmonia, respectively, due to
neglecting higher order terms in $v$. The
uncertainty of the matching coefficients between QCD and the
effective nonrelativistic theory is another
source of systematic biases. We have assumed that the coefficient
$c_1(m)$, in
front of the correction to the kinetic energy, equals 1.
As can be seen from Table~\ref{ren2t}, such coefficients can easily
differ by as much as 20~\% from this tree-level value in the case of
bottomonia and by a factor 2 for charmonia. Such an effect
on $c_1(m)$ would result in
shifts of certain bottomonia states by about 4~MeV and charmonia
states of up to 50~MeV. Higher order corrections to $c_2(m)$ and $c_3(m)$
will also have an effect but at present the value of $c_1(m)$
constitutes the dominant uncertainty.

We conclude that the deviations between experiment and prediction can
be explained in terms of quenching and higher order relativistic
corrections, i.e.\
${\mathcal O}(v^4)$-terms as well as 
${\mathcal O}(\ln m/\mu)$ and ${\mathcal O}(m/\mu\ln m/\mu)$
uncertainties in the
matching coefficients of the ${\mathcal O}(v^2)$ terms.
Inclusion of sea quarks seems to improve agreement with
experiment
but is unlikely to reduce
deviations by more than an average of 4--5~MeV per state. In the case of
bottomonia, we estimate the impact of
higher order correction terms to be about
twice as large. For
charmonia the effect of ${\mathcal O}(v^4)$ corrections might be as
large as 10--20~MeV 
while the impact of the matching constants of the ${\mathcal O}(v^2)$
terms is even larger.
Thus the agreement on a 20 MeV level appears to be somewhat
fortuitous. However, this outcome is not a complete surprise since
many effects seem to affect the
spectrum as a whole, rather than individual splittings, and can
compensate each other.

In this first glimpse at the spectrum,
we have not yet included a running coupling into the parametrization
of the potentials. The SD and MD corrections as well as the correction
to the kinetic energy have so far been treated as first order
perturbations only. We will improve on these two points
in a detailed spectroscopy study~\cite{spectro}, in which we
are going to elaborate on the effect of higher order
relativistic uncertainties on individual states in a more
systematic manner.

\section{Conclusions and outlook}
We have determined the complete ${\mathcal O}(v^2)$ relativistic
corrections to the static interquark potential in SU(3)
gauge theory.
We find reliable renormalized
potentials with good scaling behavior. As in our SU(2) study~\cite{wachter}
we report clear evidence for a
$1/r^2$ scalar exchange contribution in the long range spin-orbit
potential $V_1'$
at the level of 20~\% of the Coulomb part of the static
potential at inverse lattice spacings of $2$--$3$ GeV. The
other SD potentials are found to be short ranged and
are well understood by means of perturbation theory.
From $V_2'$, we obtain the result
$\alpha_V^{n_f=0}(\mu)=0.124\pm 0.005\pm 0.003$
at $\mu\approx 9.2$~GeV, where the first error is statistical
and the second one accounts for the differences between one- and
two-loop estimates. This value is
significantly smaller than the estimate $\alpha_V(\mu)=0.138$,
obtained from the average plaquette. 

All MD potentials contain contributions, that are linear
in the quark separation, and are
in qualitative agreement with minimal area law expectations.
The potential $\nabla^2V_a^E$, that modifies the central
force, is found to be Coulomb-like and has a significant effect
on spectroscopy since it increases the effective Coulomb force by 2~\%
in the case of bottomonia and by as much as 35--40~\% for charmonia.
A similar behavior is expected from dual QCD~\cite{dqcd}.

As an application, quarkonia spectra are determined. We are
able to reproduce the experimental levels with
an average error of $12.5$~MeV for $\Upsilon$ states and
$22$~MeV for $J/\psi$ states.
A reduction of these deviations should be achieved by
incorporating improved parametrizations of the lattice
potentials, that account 
for a weakening of the effective QCD
coupling at small separations,
into the spectroscopy. Such a refined
analysis is in progress~\cite{spectro}.
We estimate a further improvement
of up to 4~MeV per state if dynamical sea
quarks are included, while higher order relativistic corrections
and uncertainties in the perturbative matching constants
might shift certain levels by as much as 5--10~MeV for bottomonia-
and up to 50~MeV for charmonia-states. Uncertainties in the
(perturbative) matching constants between QCD and the effective
nonrelativistic theory are likely to have a larger impact than
${\mathcal O}(v^4)$ corrections, and should be reduced.
Our results are compatible with heavy quark pole masses
$m^{\mbox{\scriptsize pole}}_b=4.7(2)$~GeV and $m_c^{\mbox{\scriptsize
pole}}=1.4(2)$~GeV.

The approach presented in this article can be used to obtain
optimized wave functions for creation of a
quarkonium state with particular quantum numbers within
the complementary lattice NRQCD method~\cite{sesam}.
We intend to extend this application to discrete
finite boxes with periodic boundary conditions, in order to shape even
better basis states and to simulate finite
size effects that one might expect in lattice NRQCD studies.
From Fig.~\ref{wave}, it is obvious that on volumes with
a spatial extent of typically less than 2~fm, excited
state wave functions become squeezed, and the
corresponding energy eigenvalues might be significantly affected.

Application of the Schr\"odinger-Pauli approach to
the spectrum of $B_c$ states as well as a determination
of quarkonia decay constants is in progress.
It appears worthwhile to consider calculations on anisotropic
lattices,
to reduce systematic uncertainties on the potentials,
arising from the temporal discretization of the lattice.
In order to keep uncertainties in the perturbative matching
constants between the effective theory at a scale $\mu=\pi/a$ and
QCD small, one would like to operate at spacings $a\approx
\pi/m$ where
$m$ might be either the bottom or the charm quark mass. The
latter would require an improved lattice action.

\acknowledgements
GSB thanks the Physics Department of the University of Glasgow
for hospitality during part of this work. During his visit
he enjoyed fruitful discussions with Sara Collins and
Christine Davies. He also acknowledges inspiring discussions
with Nora Brambilla and Marshall Baker.
GSB has been supported by EU grant No.\ ERB
CHBG CT94-0665 and PPARC grant No.\ GR/K55738. We appreciate support from the
EU (grant Nos.\ SC1*-CT91-0642 and CHRX-CT92-00551) and the
Deutsche Forschungsgemeinschaft (DFG grant Nos.\ Schi 257/1-4 and Schi
257/3-2). Computations have been
performed on the Connection Machines CM-5 of the Institut f\"ur
Angewandte Informatik in Wuppertal and the GMD in St.\ Augustin.
We thank Dr.\ R.~V\"olpel for his support.

\begin{table}
\caption{Simulation parameters. The physical scale has been obtained
{}from $\sqrt{\kappa}=468$~MeV.}
\label{et1}
\begin{tabular}{ccc}
&$\beta=6.0$&$\beta=6.2$\\\hline
$V=L_{\sigma}^3L_{\tau}$&$16^4$&$32^4$\\
$a/\mbox{fm}$&0.092&0.067\\
$a^{-1}/\mbox{GeV}$&2.14&2.94\\
$L_{\sigma}/\mbox{fm}$&1.47&2.15\\
$n_{\mbox{\scriptsize conf}}$&420&116\\
\end{tabular}
\end{table}

\begin{table}
\caption{Renormalization constants for magnetic and
electric ears, compared to their tadpole estimates,
$Z_{\mbox{\scriptsize tadpole}}=1/U_{\Box}$.}\label{rent}
\begin{tabular}{ccccc}
$\beta$&$U_{\Box}$&$Z_{\mbox{\scriptsize tadpole}}$
&$Z_B$&$Z_E$\\\hline
6.0&0.593682(5)&1.6844&1.6777(2)&1.6216(4)\\
6.2&0.613631(3)&1.6296&1.6249(1)&1.5782(1)\\
\end{tabular}
\end{table}

\begin{table}
\caption{Matching constants between QCD and the
effective Hamiltonian, Eqs.~(\ref{ham})--(\ref{mdpo}),
for bottom and charm
quark masses
at $\beta=6.0$ and $\beta=6.2$.}\label{ren2t}
\begin{tabular}{ccc}
&$c_2$&$c_3$\\\hline
$m_b$, $\beta=6.0$&1.034&1.181\\
$m_b$, $\beta=6.2$&1.065&1.344\\
$m_c$, $\beta=6.0$&1.220&2.159\\
$m_c$, $\beta=6.2$&1.257&2.353\\
\end{tabular}
\end{table}

\begin{table}
\caption{Fit parameters to the static potential,
Eqs.~(\ref{eq_v0_2})--(\ref{eq_v0_3}).}
\label{et2}
\begin{tabular}{cccc}
Parameter&$\beta=6.0$&$\beta=6.2$&Average value\\\hline
$K$  &0.0479(12)&0.02536(35)& K\\
$e$  &0.324(17)&0.321(7)&0.321(6)\\
$\hat{f}$&0.042(12)&0.051(5)& $0.0082(8)/\sqrt{K}$\\
$\hat{C}_V$&0.6648(78) &0.6404(26) & ---\\
$g$  &0.301(4)&0.252(2)& ---\\
$\chi^2/N_{DF}$&0.4&0.6&---\\
\end{tabular}
\end{table}

\begin{table}
\caption{Parameter values from fits to relativistic corrections
to the static potential that are relevant for spectroscopy
[Eqs.~(\ref{eqn1}), (\ref{eqn2})].
The error in square brackets is the
systematic uncertainty. Where not stated separately, it has been
included.}
\label{et3}
\begin{tabular}{cccc}
Parameter&$\beta=6.0$&$\beta=6.2$&Average value\\\hline
$\hat{b}$  &0.150(21)[01]&0.097(13)[03]&3.36(36)K\\
$h$  &0.071(5)[11]&0.065(3)[8]&0.067(9)\\
$e-h$&0.253(6)[11]&0.256(7)[8]&0.255(10)\\
\end{tabular}
\end{table}

\begin{table}
\caption{Fit parameters for the SD potentials
$V_2'$ and $V_3$ [Eqs.~(\ref{eqn3})--(\ref{eqn4})].}
\label{et4}
\begin{tabular}{ccc}
Parameter&$\beta=6.0$&$\beta=6.2$\\\hline
$c_2$      &0.274(34)[17]&0.239(16)[58]\\
$\hat{f}_2$&0.023(32)[07]&0.061(15)[05]\\
$g_2$      &0.138(40)[01]&0.133(19)[06]\\
$\chi_2^2/N_{DF}$&1.0&1.1\\\hline
$c_3$      &0.253(15)[33]&0.230(45)[28]\\
$\hat{f}_3$&0.054(11)[41]&0.047(36)[36]\\
$g_3$      &0.165(21)[34]&0.171(07)[29]\\
$\chi_3^2/N_{DF}$&0.4&1.0\\
\end{tabular}
\end{table}

\begin{table}
\caption{Fit parameters for the MD potentials
$V_b$, $V_c$ and $V_d$ [Eqs.~(\ref{eqn5}), (\ref{eqn6})].}
\label{et5}
\begin{tabular}{ccc}
Parameter&$\beta=6.0$&$\beta=6.2$\\\hline
$\hat{C}_b$      &-0.0824(71)[2]&-0.0681(38)[1]\\
$g_b$      &0.196(49)[2]&0.156(41)[1]\\
$\chi_b^2/N_{DF}$&3.0&5.9\\\hline
$g_c$      &0.304(35)[4]&0.219(12)[2]\\
$\chi_c^2/N_{DF}$&2.4&1.9\\\hline
$\hat{C}_d$      &-0.116(4)[37]&-0.122(2)[32]\\
$\chi_d^2/N_{DF}$&0.5&1.2\\
\end{tabular}
\end{table}

\begin{table}
\caption{The constant $c=C_F\alpha_s$
from the weak coupling analysis.}
\label{et6}
\begin{tabular}{cccc}
Potential&$\beta$&$c$&$\chi^2/N_{DF}$\\\hline
$V_c$ &6.0&0.58(9)&4.1\\
      &6.2&0.48(4)&5.2\\
$V_2'$&6.0&0.251(19)&0.84\\
      &6.2&0.238(12)&1.30\\
$V_3$ &6.0&0.192(22)&0.34\\
      &6.2&0.182(15)&0.46\\
$V_4$ &6.0&0.217(17)&10.3\\
      &6.2&0.199(13)&12.0\\
\end{tabular}
\end{table}

\begin{table}
\caption{$\Lambda$-parameters
from the one-loop running coupling analysis.}
\label{et7}
\begin{tabular}{cccc}
Potential&$\beta$&$\Lambda/\sqrt{\kappa}$&$\chi^2/N_{DF}$\\\hline
$V_2'$&6.0&$0.18_{-4}^{+5}$&1.25\\
      &6.2&$0.20_{-3}^{+4}$&0.93\\
$V_3$ &6.0&$0.090^{+45}_{-36}$&0.59\\
      &6.2&$0.091^{+30}_{-25}$&1.15\\
$V_4$ &6.0&$0.087^{+23}_{-21}$&4.5\\
      &6.2&$0.056^{+16}_{-13}$&12.7\\
\end{tabular}
\end{table}

\begin{table}
\caption{$\Lambda$-parameters
from the two-loop running coupling analysis.}
\label{et8}
\begin{tabular}{cccc}
Potential&$\beta$&$\Lambda/\sqrt{\kappa}$&$\chi^2/N_{DF}$\\\hline
$V_2'$&6.0&$0.41^{+9}_{-9}$&1.32\\
      &6.2&$0.45_{-7}^{+8}$&0.96\\
$V_3$ &6.0&$0.22_{-6}^{+9}$&0.59\\
      &6.2&$0.22_{-5}^{+7}$&1.15\\
$V_4$ &6.0&$0.19_{-4}^{+5}$&5.5\\
      &6.2&$0.13_{-3}^{+3}$&14.2\\
\end{tabular}
\end{table}

\begin{table}
\caption{The bottomonium spectrum.}
\label{spect1t}
\begin{tabular}{ccccc}
$n^{(2S+1)}L_J$&$\beta=6.0$&$\beta=6.2$&$e=0.40$&experiment\\\hline
$1^1S_0$ &  9.477&  9.476&  9.415&\\
$1^3S_1$ &  9.521&  9.526&  9.504&9.460\\
$2^1S_0$ &  9.980&  9.980&  9.961&\\
$2^3S_1$ & 10.007& 10.010& 10.008&10.023\\
$3^1S_0$ & 10.328& 10.328& 10.311&\\
$3^3S_1$ & 10.351& 10.354& 10.348&10.355\\
$4^1S_0$ & 10.619& 10.619& 10.597&\\
$4^3S_1$ & 10.640& 10.642& 10.630&10.580\\
$1^1P_1$ &  9.879&  9.879&  9.889&\\
$1^3P_0$ &  9.866&  9.866&  9.867&9.860\\
$1^3P_1$ &  9.878&  9.878&  9.886&9.892\\
$1^3P_2$ &  9.882&  9.883&  9.895&9.913\\
$2^1P_1$ & 10.238& 10.238& 10.243&\\
$2^3P_0$ & 10.226& 10.225& 10.223&10.232\\
$2^3P_1$ & 10.237& 10.237& 10.240&10.255\\
$2^3P_2$ & 10.241& 10.242& 10.249&10.269\\
$1^1D_2$ & 10.120& 10.121& 10.136&\\
$1^3D_1$ & 10.121& 10.121& 10.134&\\
$1^3D_2$ & 10.121& 10.122& 10.137&\\
$1^3D_3$ & 10.119& 10.120& 10.137&\\
\end{tabular}
\end{table}

\begin{table}
\caption{The charmonium spectrum. Only the states that are
identified by an asterisk lie below the $D\overline{D}$ threshold and have
been used to fix the scale $\kappa$ and quark mass $m_c$.}
\label{spect2t}
\begin{tabular}{ccccc}
$n^{(2S+1)}L_J$&$\beta=6.0$&$\beta=6.2$&$e=0.40$&experiment\\\hline
$1^1S_0$ &  3.010&  3.001&  2.966&2.980*\\
$1^3S_1$ &  3.134&  3.143&  3.175&3.097*\\
$2^1S_0$ &  3.591&  3.582&  3.560&3.594*\\
$2^3S_1$ &  3.685&  3.688&  3.705&3.686*\\
$3^1S_0$ &  4.017&  4.004&  3.978&\\
$3^3S_1$ &  4.102&  4.098&  4.106&4.040\\
$4^1S_0$ &  4.371&  4.354&  4.324&\\
$4^3S_1$ &  4.452&  4.442&  4.442&4.415\\
$1^1P_1$ &  3.468&  3.472&  3.486&\\
$1^3P_0$ &  3.452&  3.451&  3.442&3.415*\\
$1^3P_1$ &  3.479&  3.482&  3.490&3.511*\\
$1^3P_2$ &  3.465&  3.471&  3.491&3.556*\\
$2^1P_1$ &  3.915&  3.913&  3.916&\\
$2^3P_0$ &  3.893&  3.886&  3.870&\\
$2^3P_1$ &  3.922&  3.919&  3.917&\\
$2^3P_2$ &  3.916&  3.914&  3.924&\\
$1^1D_2$ &  3.764&  3.767&  3.782&3.770\\
$1^3D_1$ &  3.791&  3.790&  3.796&\\
$1^3D_2$ &  3.777&  3.778&  3.792&\\
$1^3D_3$ &  3.744&  3.748&  3.770&\\
\end{tabular}
\end{table}

\begin{table}
\caption{Average velocities and quark separations in bottomonia and
charmonia. The values correspond to the parameter
choice $e=0.40$, which has been used
to model the effect of dynamical sea quarks.}
\label{veloc}
\begin{tabular}{ccccc}
$nL$&$\langle v_b^2\rangle$&$\langle v_c^2\rangle$&
$\sqrt{\langle r_b^2\rangle}/$fm&$\sqrt{\langle r_c^2\rangle}/$fm\\\hline
$1S$&0.080&0.27&0.24&0.43\\
$2S$&0.081&0.35&0.51&0.85\\
$3S$&0.096&0.44&0.73&1.18\\
$4S$&0.112&0.52&0.93&1.47\\
$1P$&0.068&0.29&0.41&0.67\\
$2P$&0.085&0.39&0.65&1.04\\
$1D$&0.075&0.34&0.54&0.87\\
\end{tabular}
\end{table}

\begin{figure}
\unitlength 1cm
\begin{center}
\begin{picture}(12,9.5)
\put(0,0){\epsfxsize=12cm\epsfbox{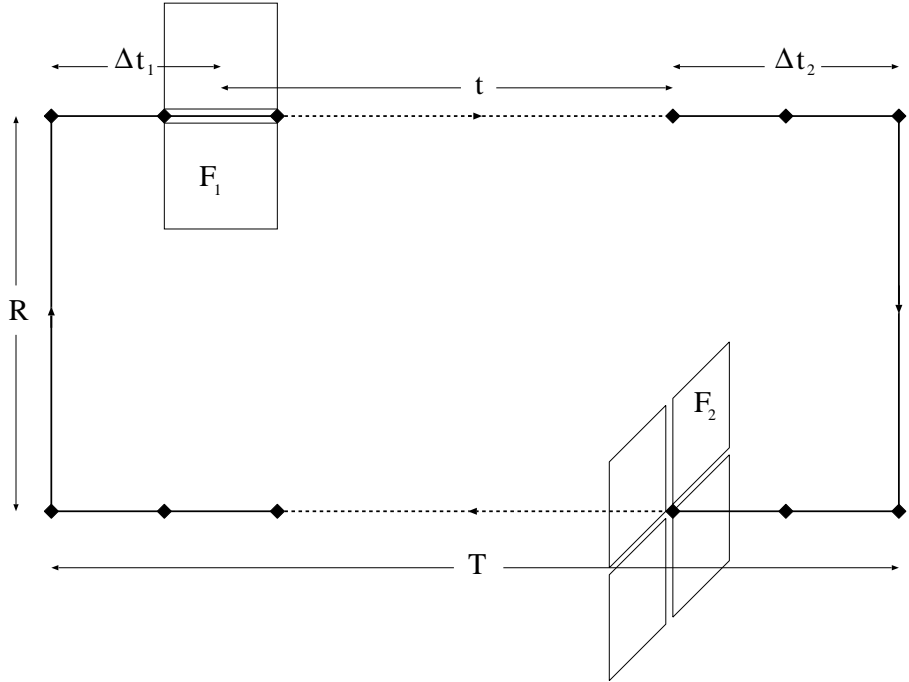}}
\end{picture}
\end{center}
\caption{Lattice definition of the nominator of Eq.~(\ref{dex}) for
the example of
$F_1$ being an electric ear and $F_2$ being a magnetic ear.}
\label{ohr}
\end{figure}

\begin{figure}
\begin{center}
{\epsfxsize=12.5cm\epsfbox{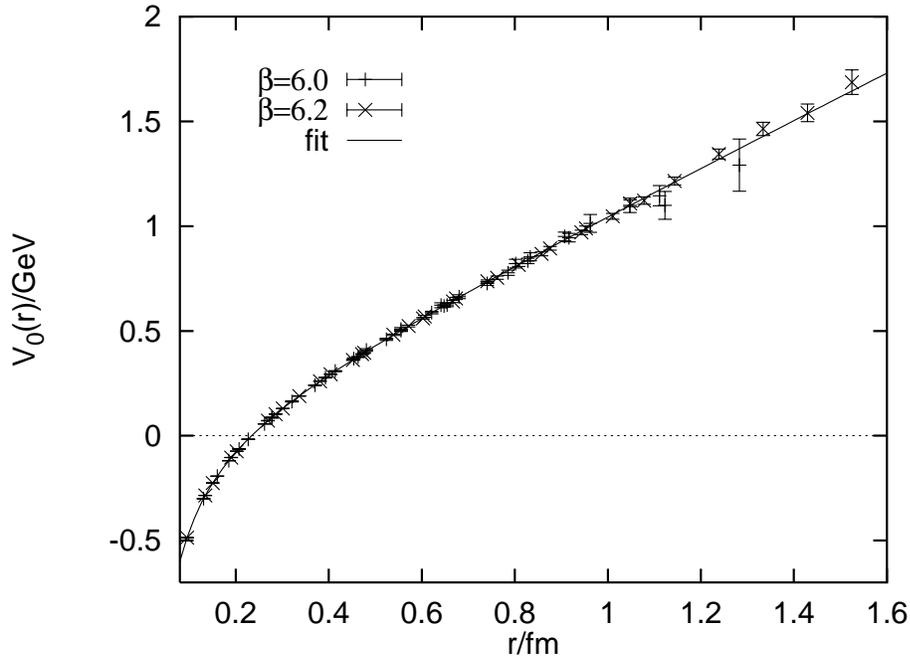}}
\end{center}
\caption{Corrected static potential
$V_{0,\mbox{\scriptsize cont}}$
at $\beta=6.0$ and $\beta=6.2$.
The fit curve corresponds to the parametrization Eq.~(\ref{eq_v0_3})
with the parameter values listed in the last column of Table~\ref{et2}.}
\label{v0}
\end{figure}

\begin{figure}
\begin{center}

{\epsfxsize=13cm\epsfbox{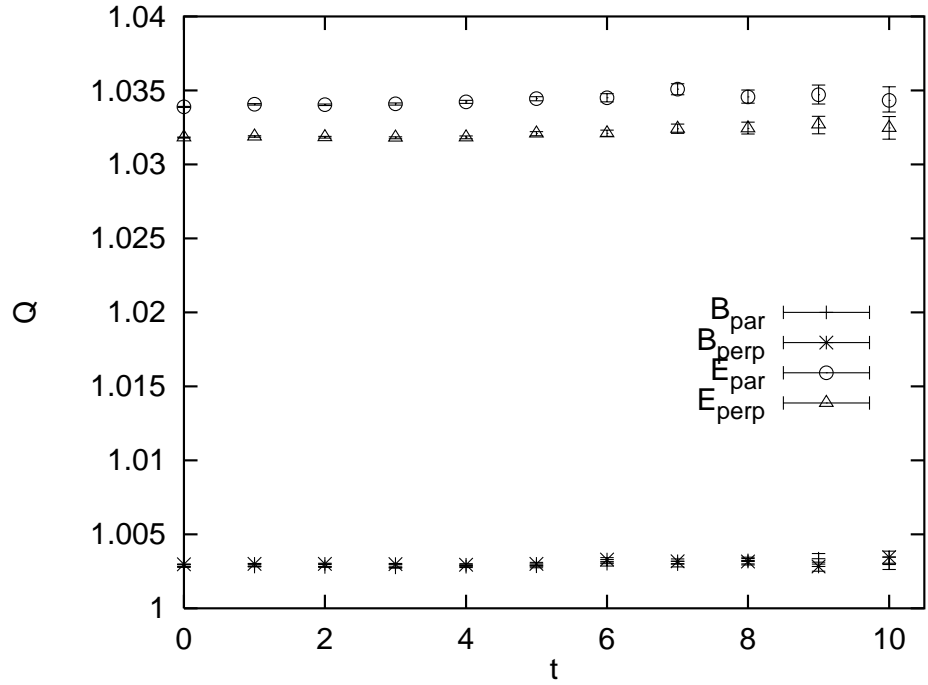}}

\end{center}
\caption{The ratio of tadpole and HM renormalization
constants $Q$ [Eq.~(\ref{qdef})]
as a function of $t$ at $R=4$, $\beta=6.2$.}
\label{renorm1}
\end{figure}

\begin{figure}
\begin{center}

{\epsfxsize=13cm\epsfbox{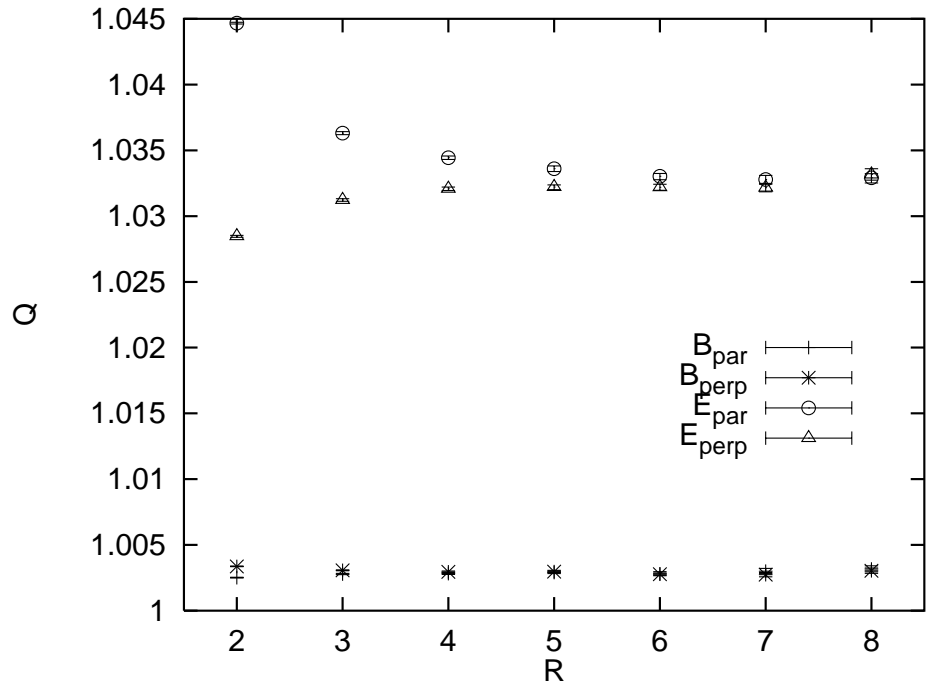}}

\end{center}
\caption{The ratio $Q$ as a function of
$R$ at $\beta=6.2$ (for large $t$).}
\label{renorm2}
\end{figure}

\begin{figure}
\begin{center}

{\epsfxsize=13cm\epsfbox{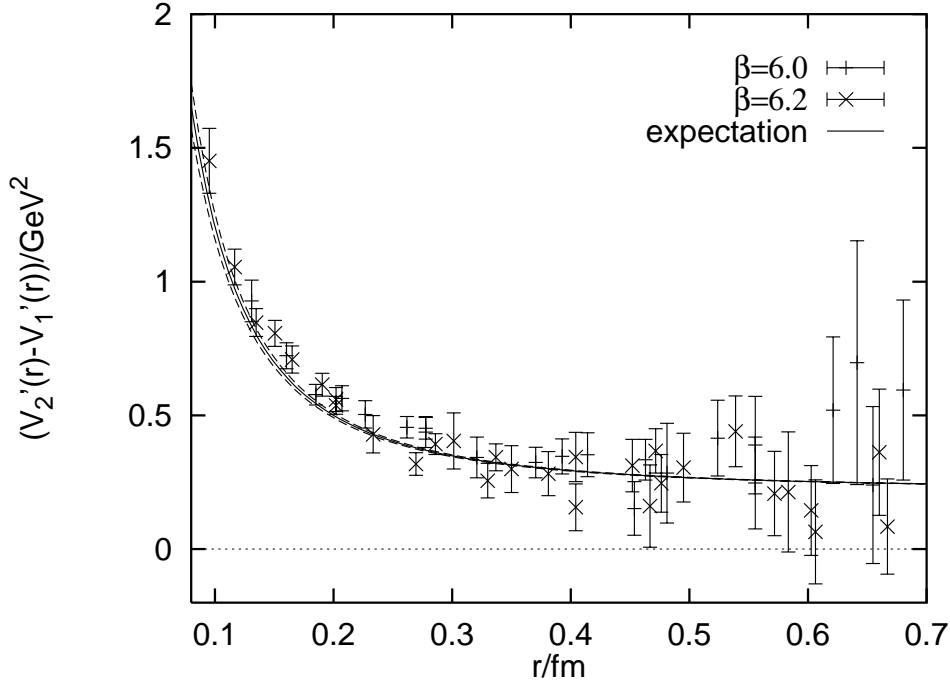}}

\end{center}
\caption{Test of the Gromes relation, Eq.~(\ref{grom}). The
combination
$V_2'-V_1'$ is compared to the static force as obtained from the
parametrization Eq.~(\ref{eq_v0_3}).}
\label{v12}
\end{figure}

\begin{figure}
\begin{center}

{\epsfxsize=13cm\epsfbox{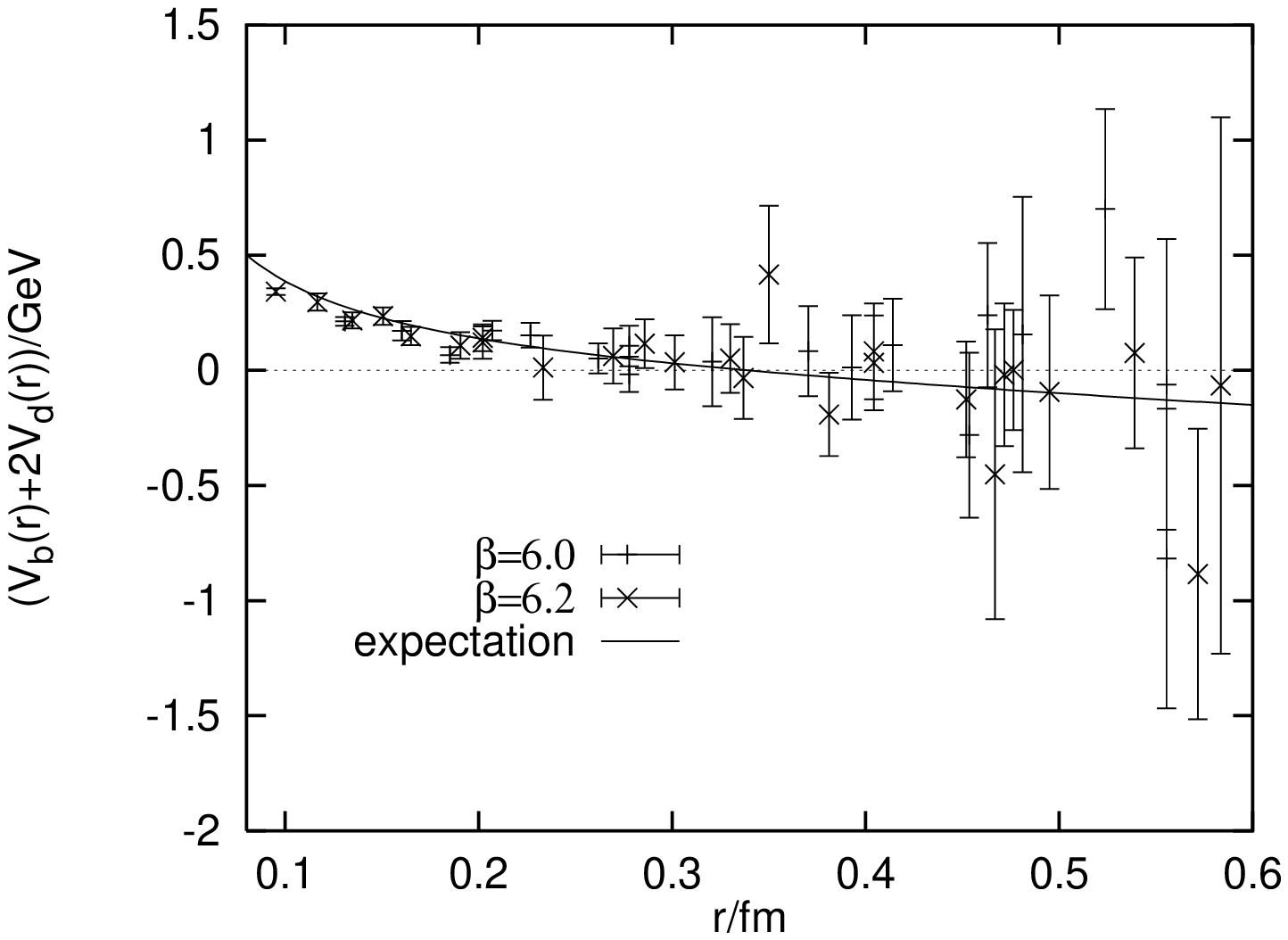}}

\end{center}
\caption{Test of the BBMP relation, Eq.~(\ref{bram2}). The
combination
$V_b+2V_d$ is compared to its expectation
as obtained from the
parametrization Eq.~(\ref{eq_v0_3})
of the static potential.}
\label{vbd}
\end{figure}

\begin{figure}
\begin{center}

{\epsfxsize=13cm\epsfbox{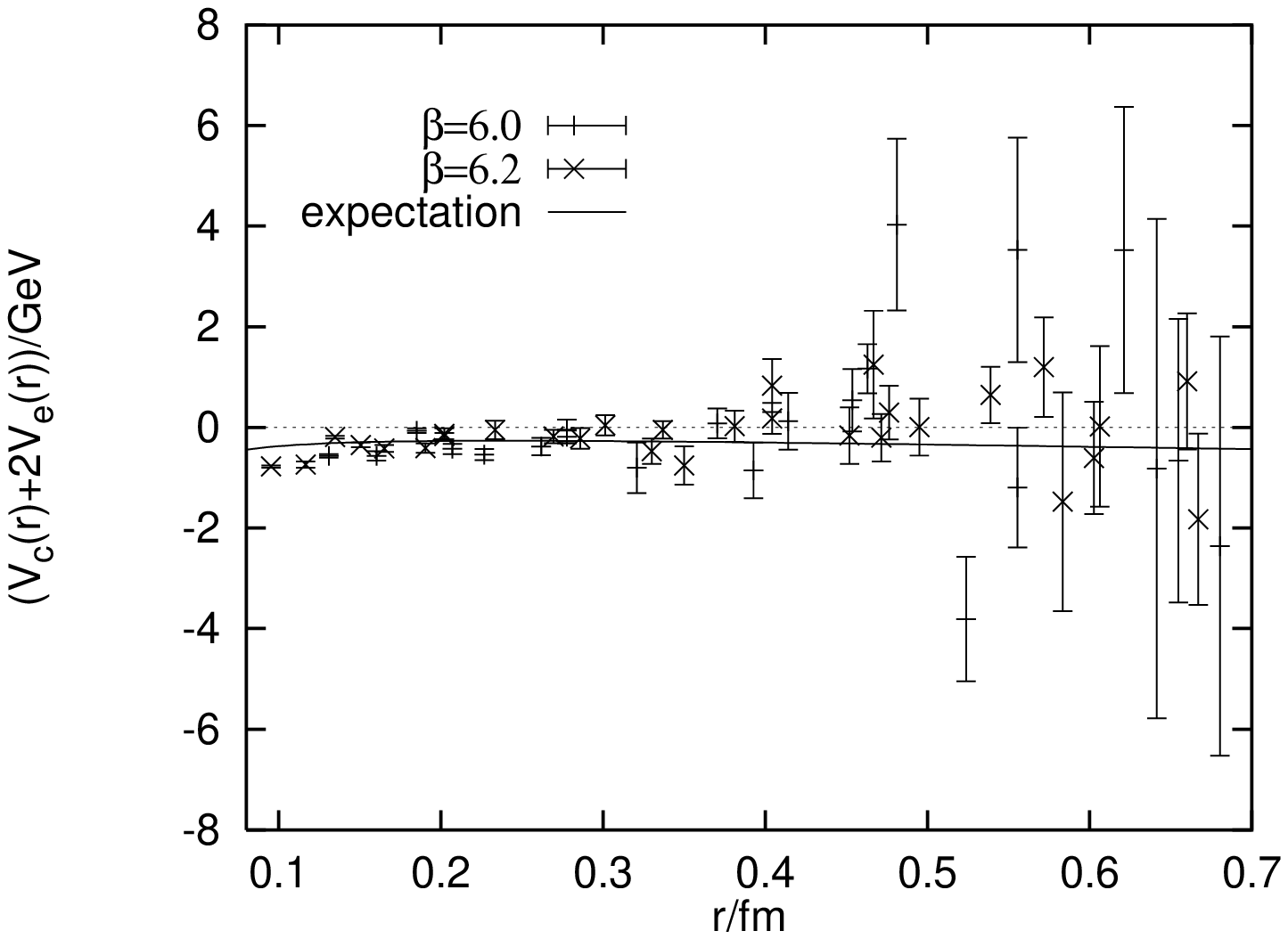}}

\end{center}
\caption{Test of the BBMP relation, Eq.~(\ref{bram}). The
combination
$V_c+2V_e$ is compared to its expectation
as obtained from the
parametrization Eq.~(\ref{eq_v0_3})
of the static potential.}
\label{vce}
\end{figure}

\begin{figure}
\begin{center}

{\epsfxsize=13cm\epsfbox{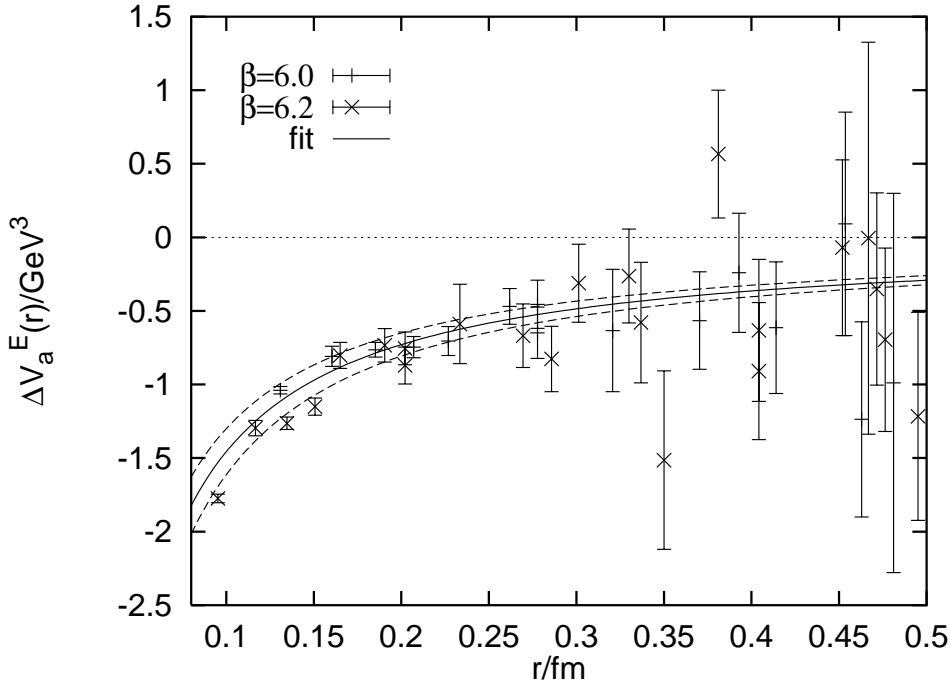}}

\end{center}
\caption{The potential $\nabla^2V_a^E$, together with
a fit curve of the form $\nabla^2V_a^E(r)=-b/r$,
with $b=(0.86\pm 0.05\mbox{~GeV})^2$. The constants
$C_a^E$ have been subtracted from the data points.}
\label{vae}
\end{figure}

\begin{figure}
\begin{center}

{\epsfxsize=13cm\epsfbox{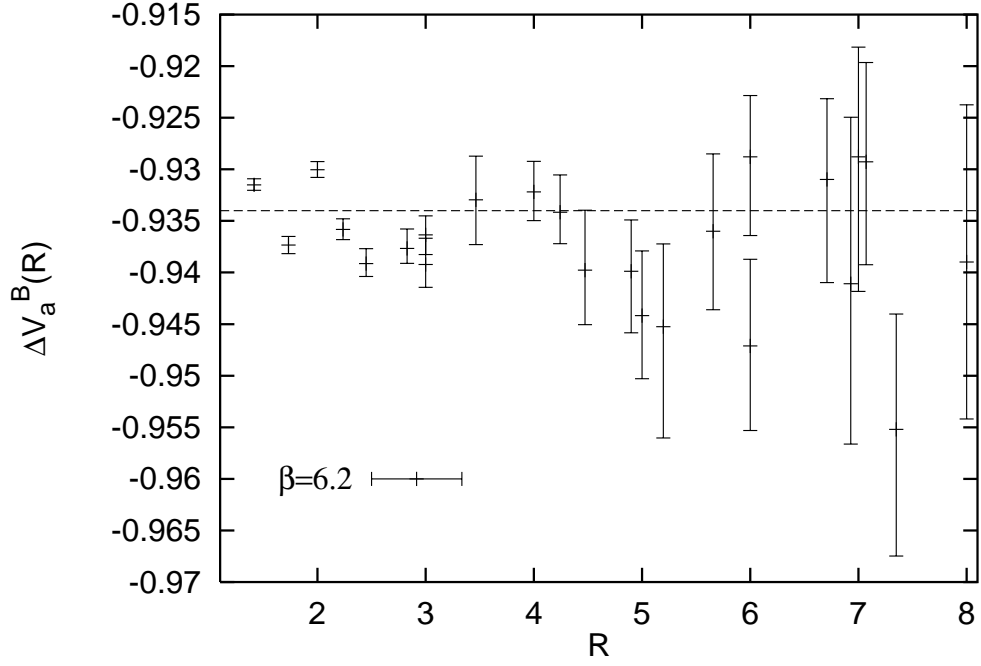}}

\end{center}
\caption{The potential $\nabla^2V_a^B$ at $\beta=6.2$
in lattice units (statistical errors only).}
\label{vab}
\end{figure}

\begin{figure}
\begin{center}

{\epsfxsize=13cm\epsfbox{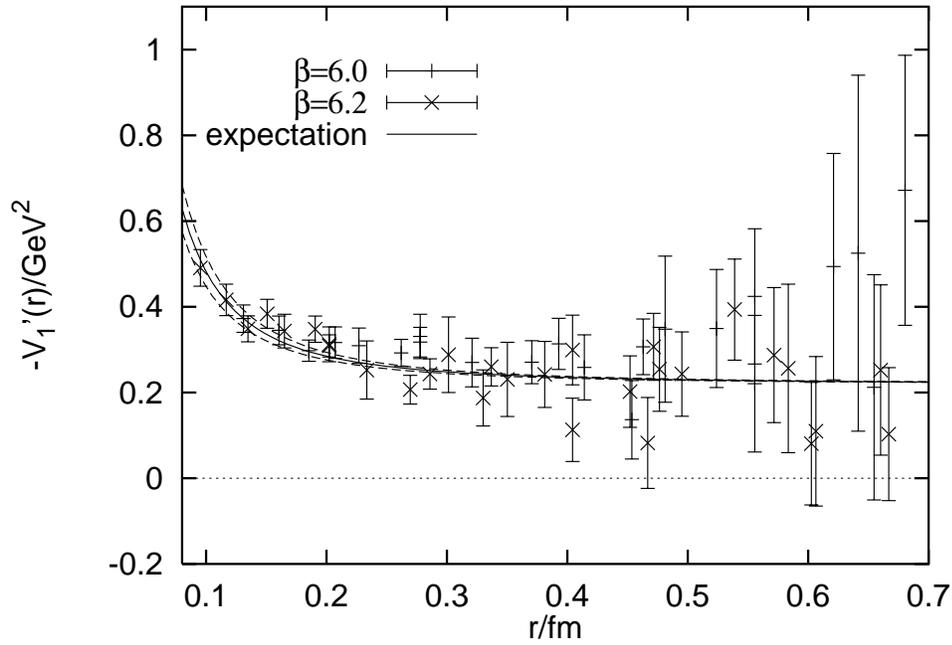}}

\end{center}
\caption{The spin-orbit potential $V_1'$, together with
a fit curve of the form $-V_1'(r)=\kappa+h/r^2$. with $h=0.067(9)$.}
\label{v1}
\end{figure}

\begin{figure}
\begin{center}

{\epsfxsize=13cm\epsfbox{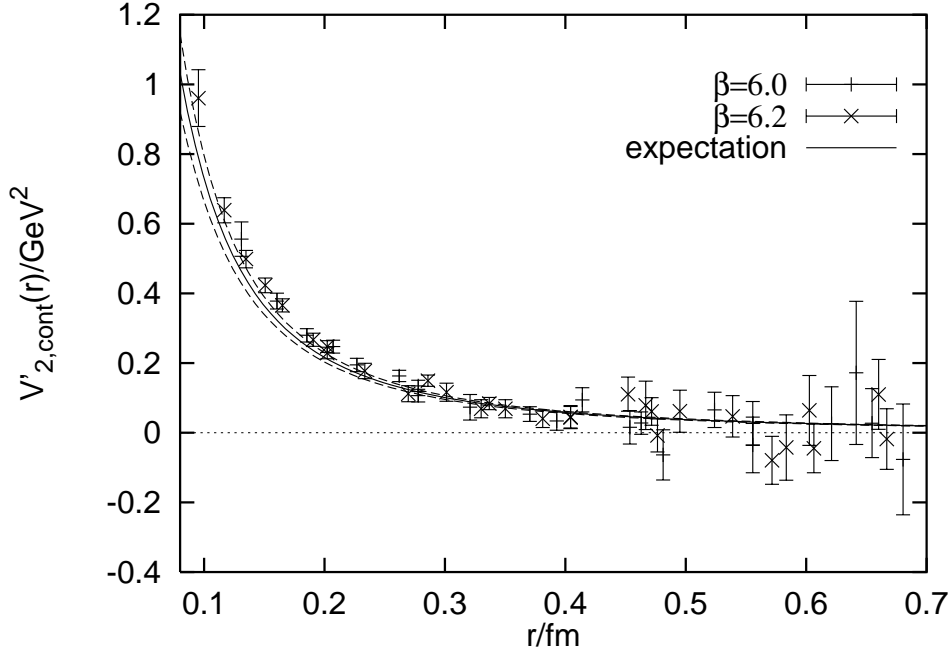}}

\end{center}
\caption{The spin-orbit potential
$V_{2,\mbox{\scriptsize cont}}'$
in comparison to the continuum expectation from
Eq.~(\ref{v2exp}).}
\label{v2}
\end{figure}

\begin{figure}
\begin{center}

{\epsfxsize=13cm\epsfbox{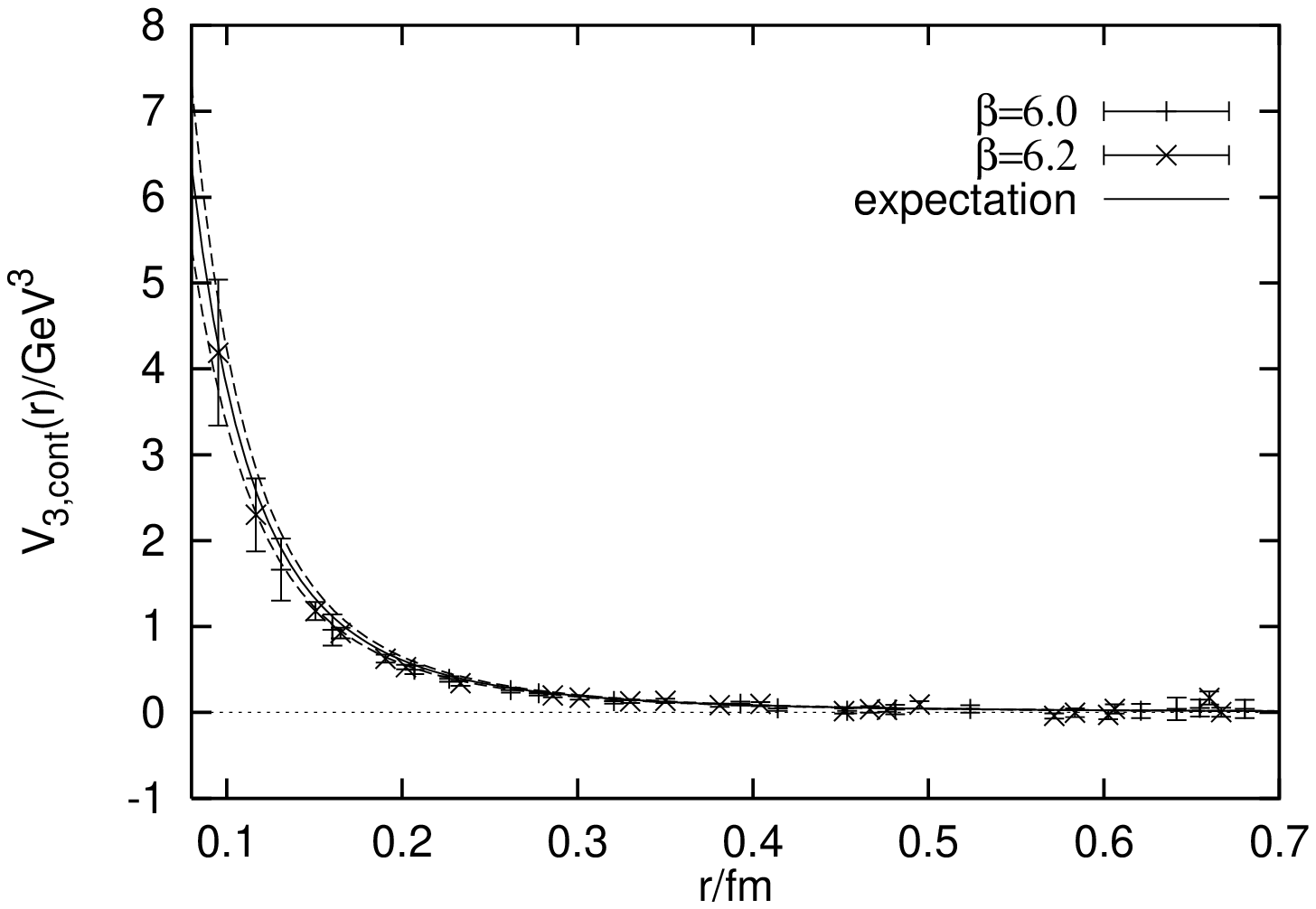}}

\end{center}
\caption{The spin-spin potential
$V_{3,\mbox{\scriptsize cont}}$
in comparison to the continuum expectation from
Eq.~(\ref{v3exp}).}
\label{v3}
\end{figure}

\begin{figure}
\begin{center}

{\epsfxsize=13cm\epsfbox{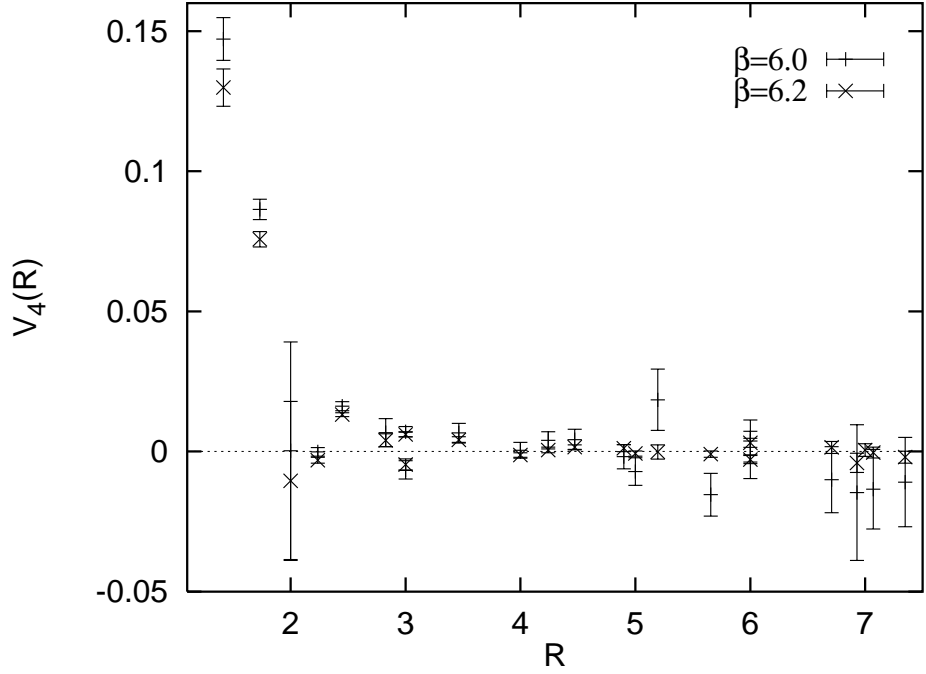}}

\end{center}
\caption{The spin-spin potential $\tilde{V}_4$ for the two $\beta$-values in
lattice units.}
\label{v4}
\end{figure}

\begin{figure}
\begin{center}

{\epsfxsize=13cm\epsfbox{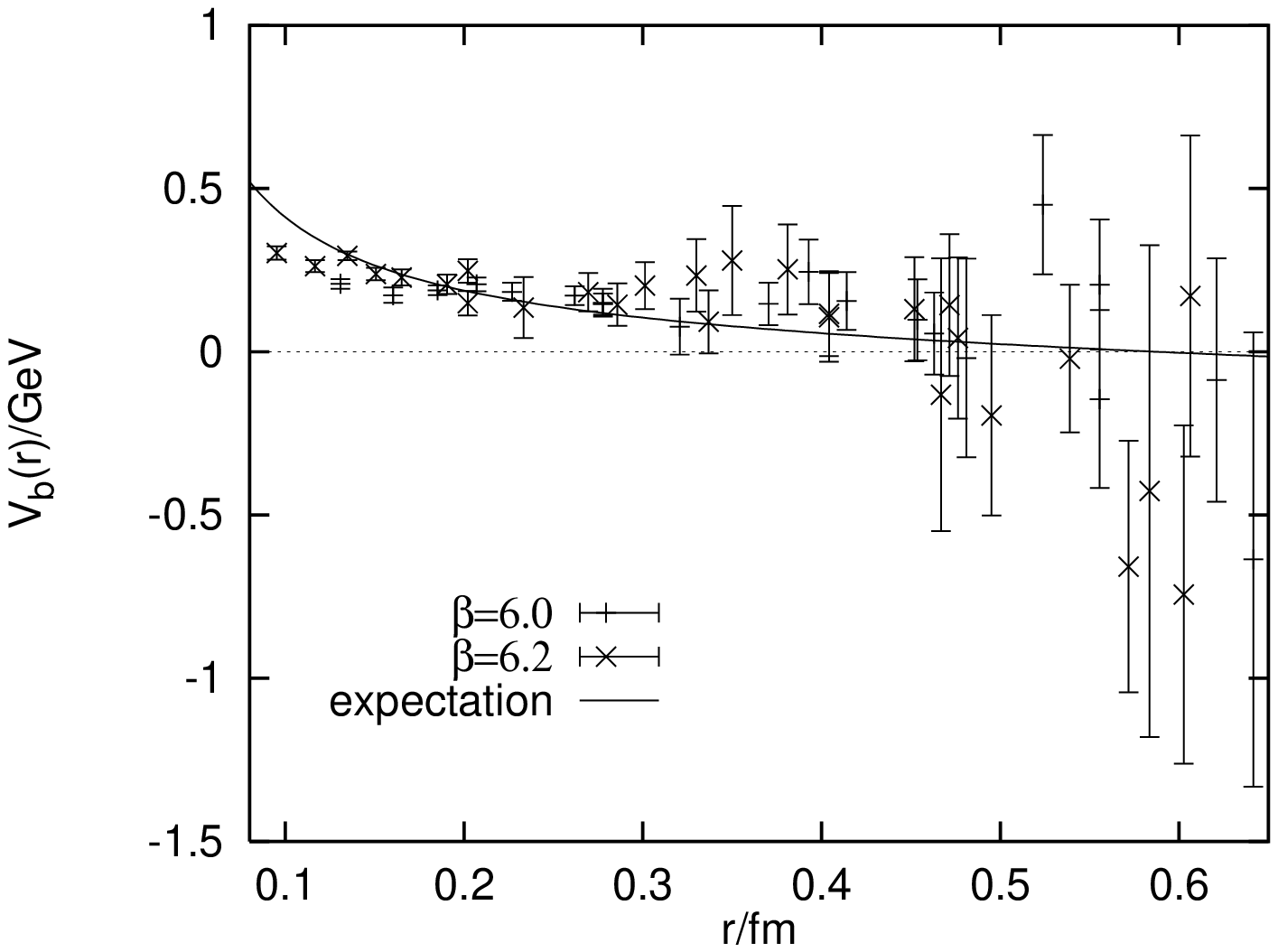}}

\end{center}
\caption{The MD potential $V_{b,\mbox{\scriptsize cont}}$ in
comparison to the continuum expectation from
Eq.~(\ref{vbpa}).}
\label{vb}
\end{figure}

\begin{figure}
\begin{center}

{\epsfxsize=13cm\epsfbox{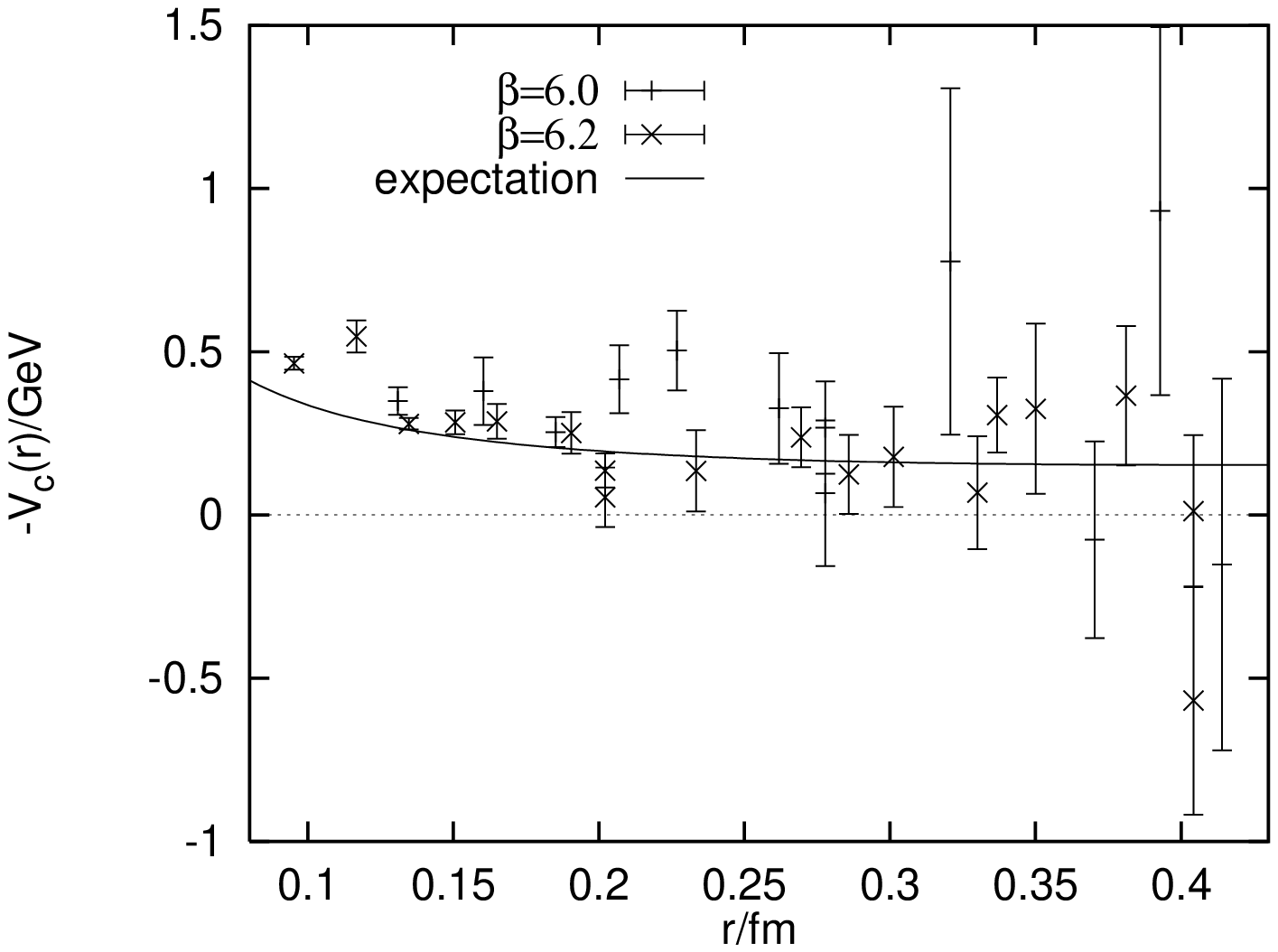}}

\end{center}
\caption{The MD potential $-V_{c,\mbox{\scriptsize
cont}}$ in comparison to the continuum expectation from
Eq.~(\ref{vbpa}).}
\label{vc}
\end{figure}

\begin{figure}
\begin{center}

{\epsfxsize=13cm\epsfbox{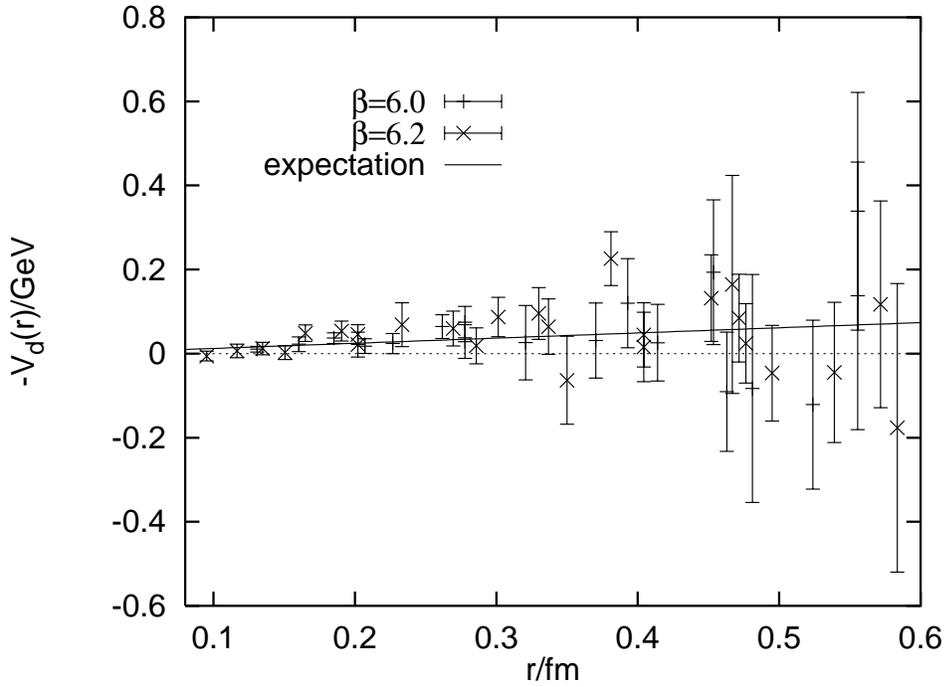}}

\end{center}
\caption{The MD potential $-V_d$ in
comparison to the continuum expectation from
Eq.~(\ref{mdpa}). The constants $C_d$ have been subtracted from the
data points.}
\label{vd}
\end{figure}

\begin{figure}
\begin{center}

{\epsfxsize=13cm\epsfbox{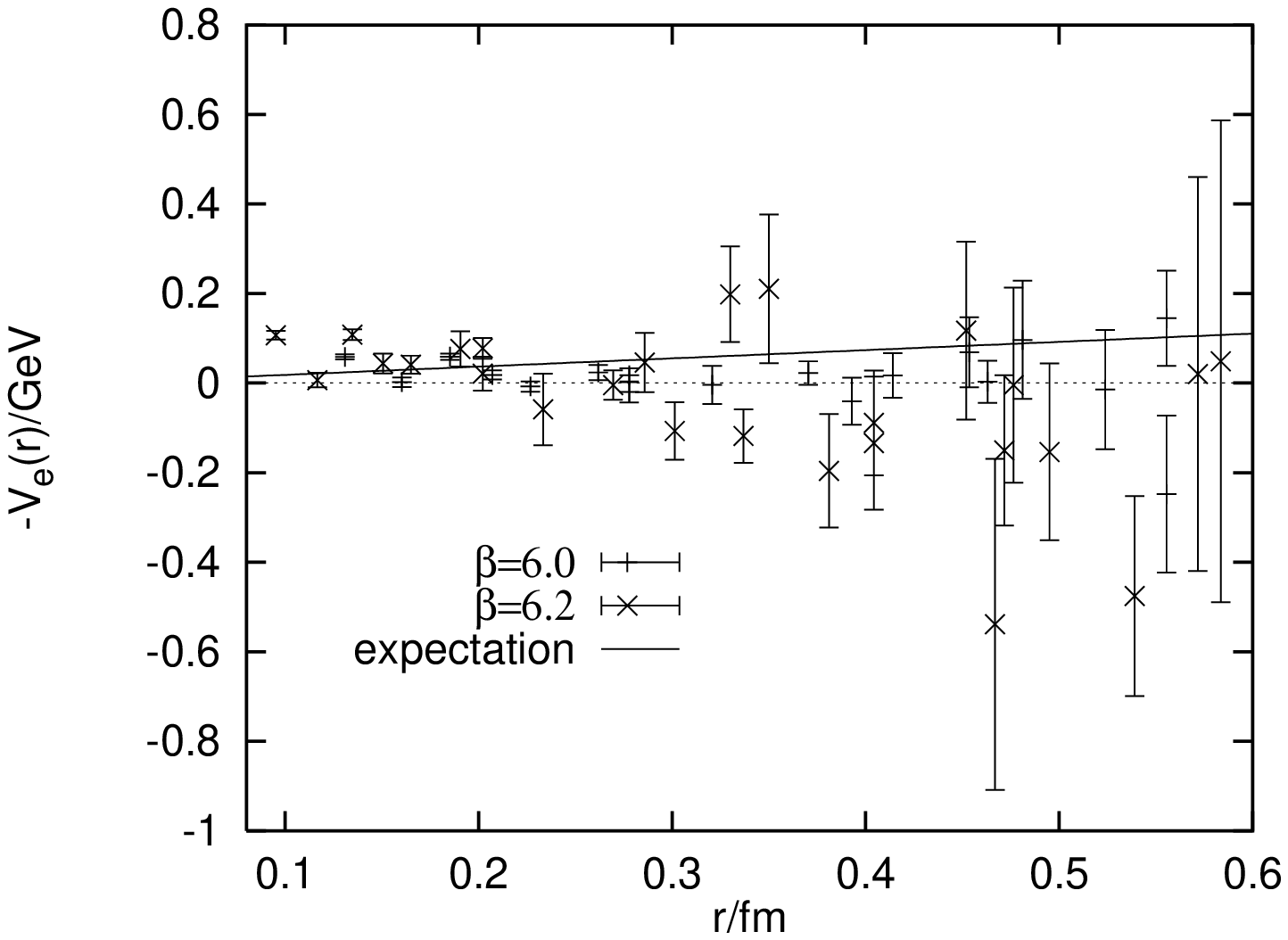}}

\end{center}
\caption{The MD potential $-V_e$ in
comparison to the continuum expectation from
Eq.~(\ref{mdpa}).}
\label{ve}
\end{figure}

\begin{figure}
\begin{center}

{\epsfxsize=13cm\epsfbox{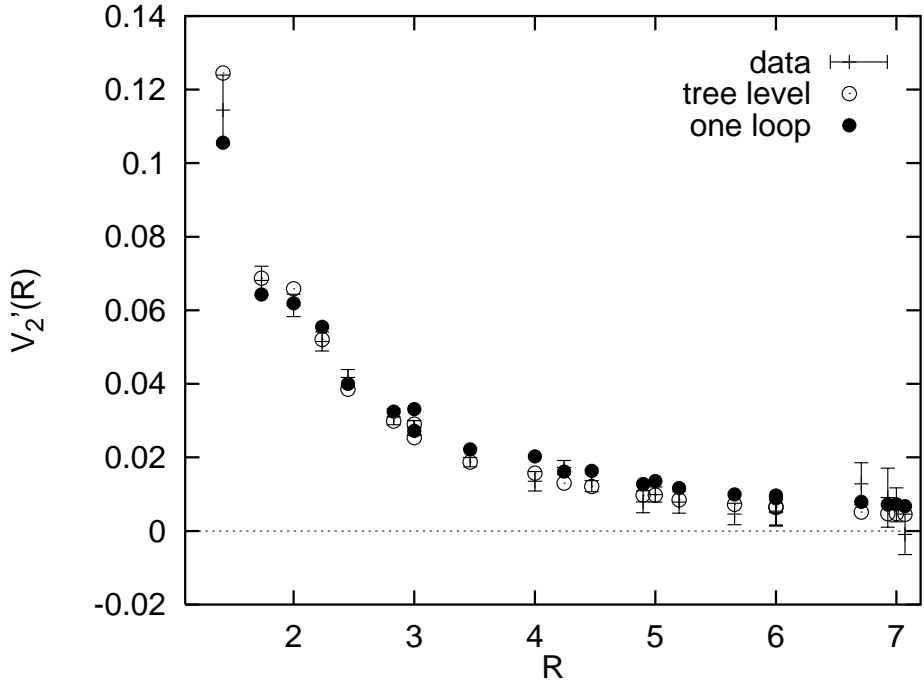}}

\end{center}
\caption{Comparison of the lattice potential $\tilde{V}_2'$
at $\beta=6.2$
to tree-level lattice perturbation theory [Eq.~(\ref{v2p})] and to
the one-loop model of Eq.~(\ref{running}).}
\label{latt_v2}
\end{figure}

\begin{figure}
\begin{center}

{\epsfxsize=13cm\epsfbox{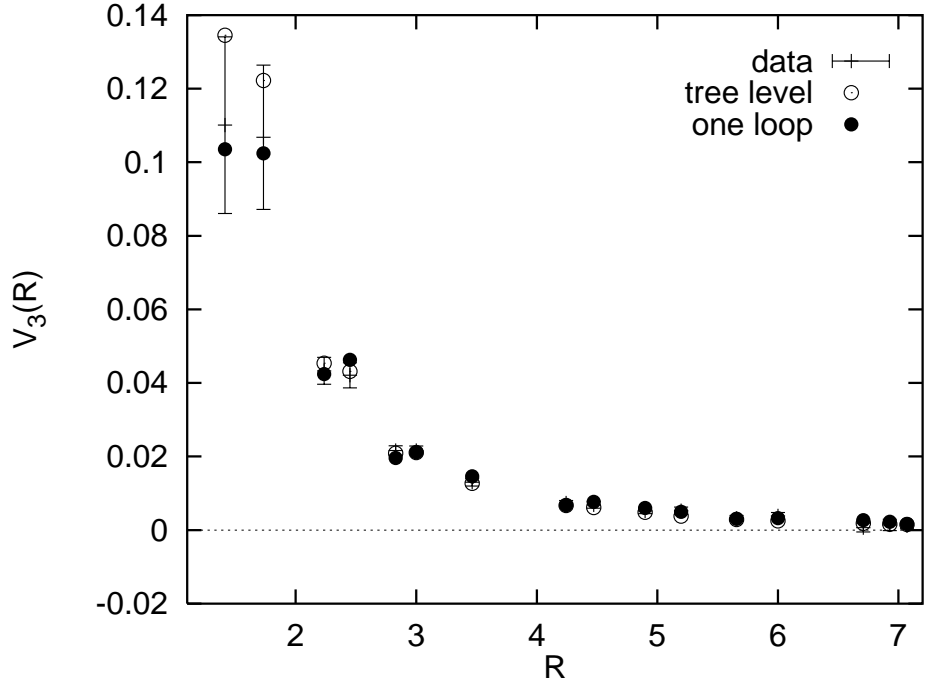}}

\end{center}
\caption{Same as Fig.~\ref{latt_v2} for $\tilde{V}_3$
[Eqs.~(\ref{latv31}) and (\ref{latv32})].}
\label{latt_v3}
\end{figure}

\begin{figure}
\begin{center}

{\epsfxsize=13cm\epsfbox{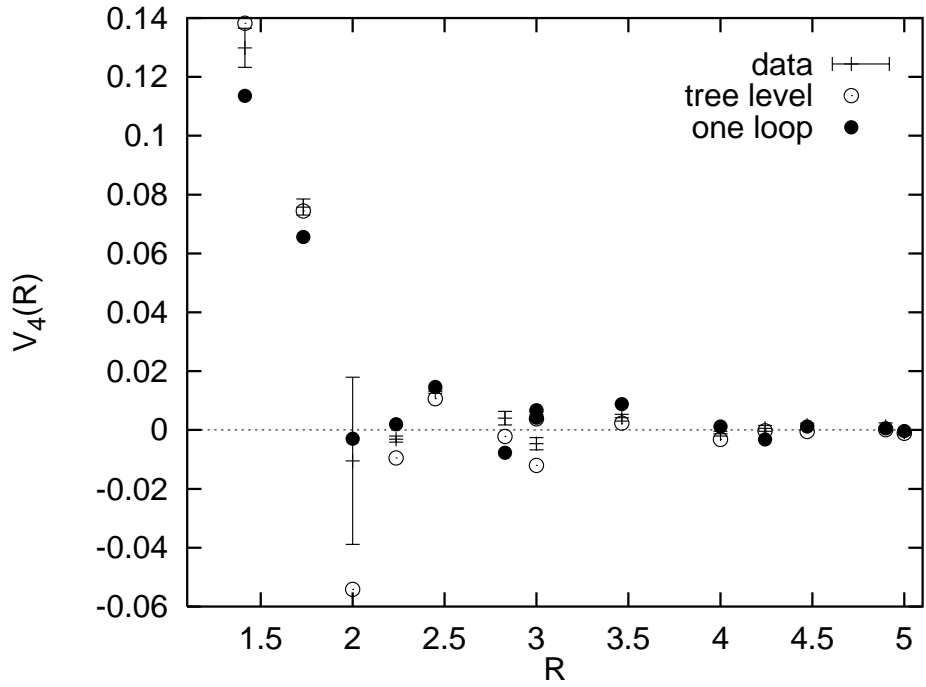}}

\end{center}
\caption{Same as Fig.~\ref{latt_v2} for $\tilde{V}_4$ [Eq.~(\ref{latv4})].}
\label{latt_v4}
\end{figure}

\begin{figure}
\begin{center}

{\epsfxsize=13cm\epsfbox{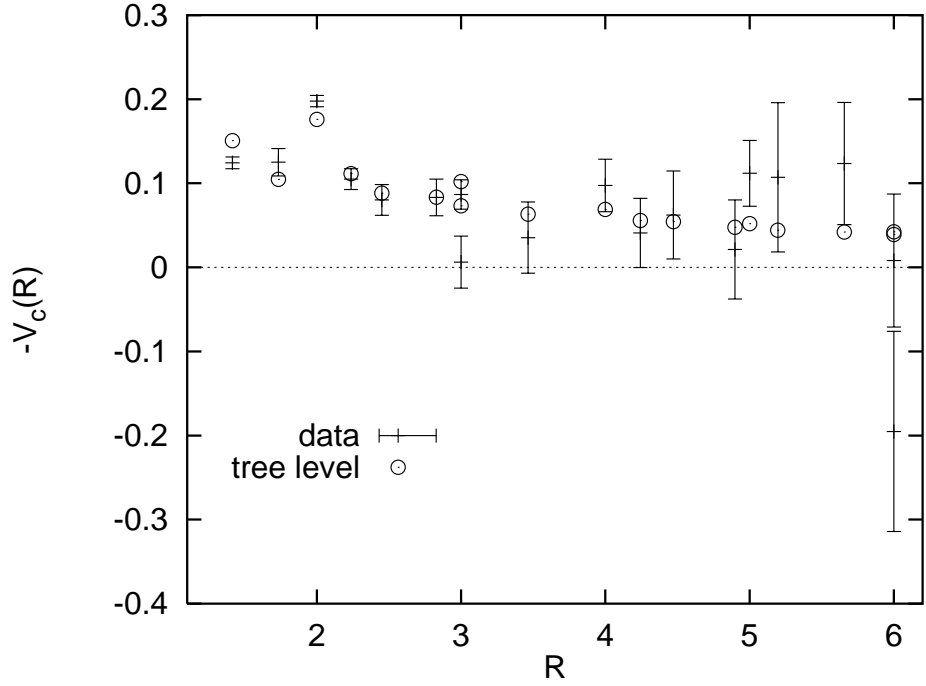}}

\end{center}
\caption{Comparison of the lattice potential $-\hat{V}_c$
at $\beta=6.2$
to tree-level lattice perturbation theory, Eq.~(\ref{latvc}).}
\label{latt_vc}
\end{figure}

\begin{figure}
\begin{center}
{\epsfysize=20cm\epsfbox{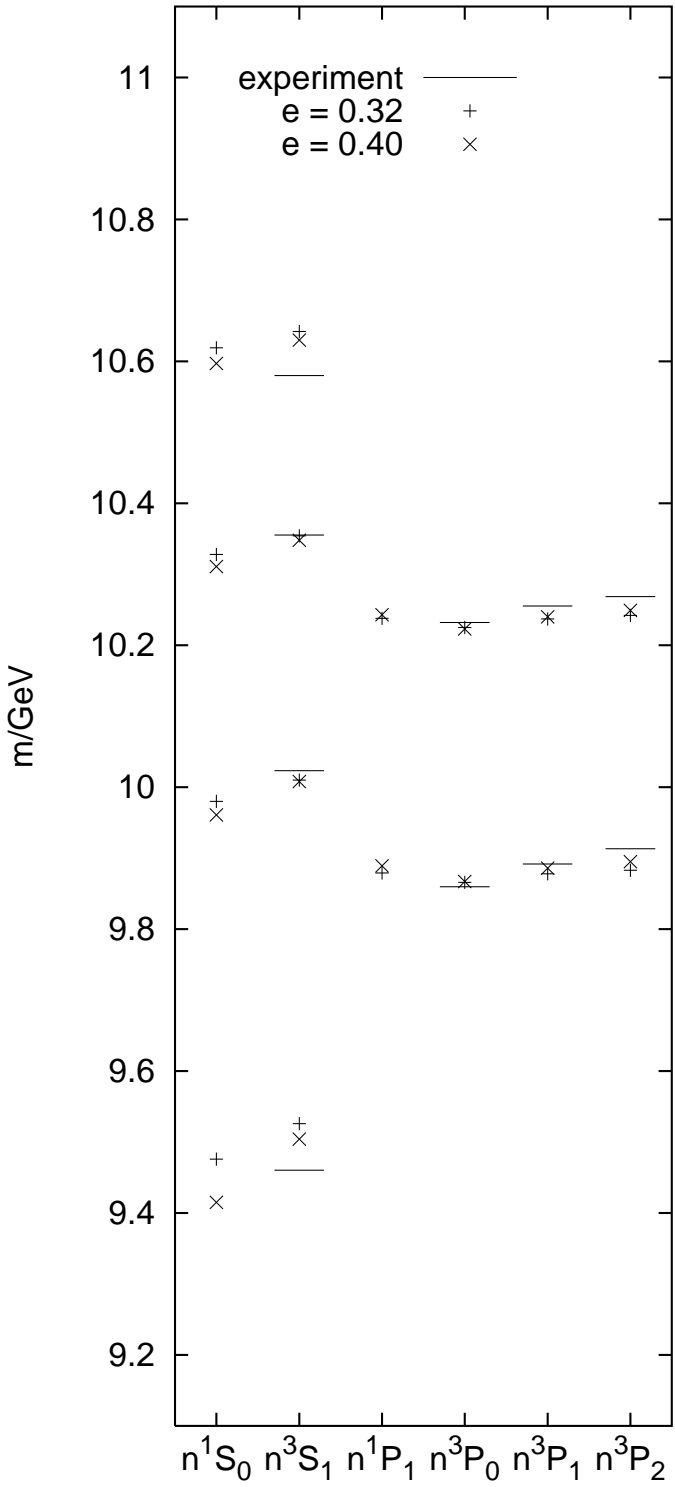}}

\end{center}
\caption{The bottomonium spectrum. The $e=0.32$ results are from the
$\beta=6.2$ analysis.}
\label{bottom}
\end{figure}

\begin{figure}
\begin{center}
{\epsfysize=20cm\epsfbox{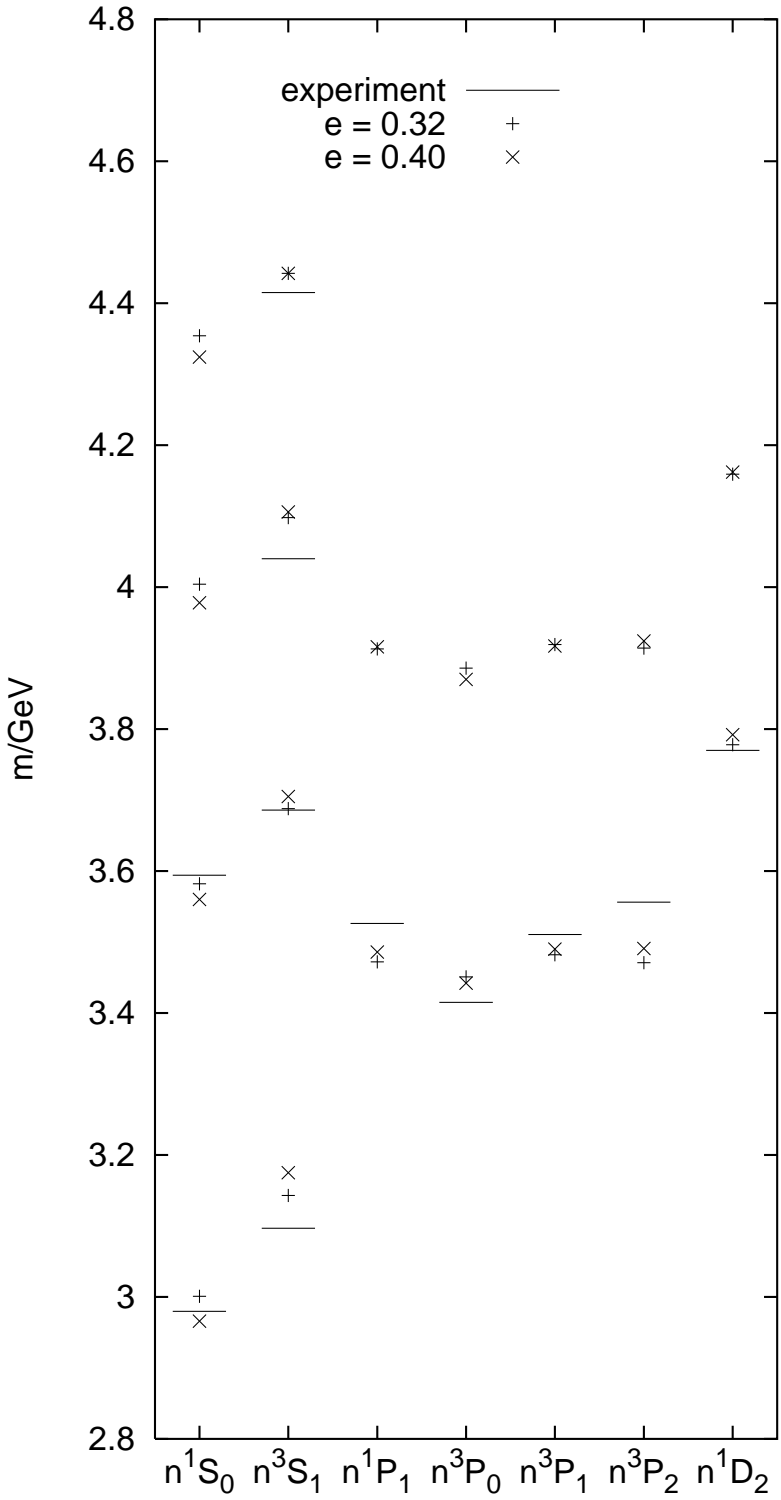}}

\end{center}

\caption{The charmonium spectrum.
The $e=0.32$ results are from the
$\beta=6.2$ analysis.}\label{charm}
\end{figure}

\begin{figure}
\begin{center}
{\epsfysize=15cm\epsfbox{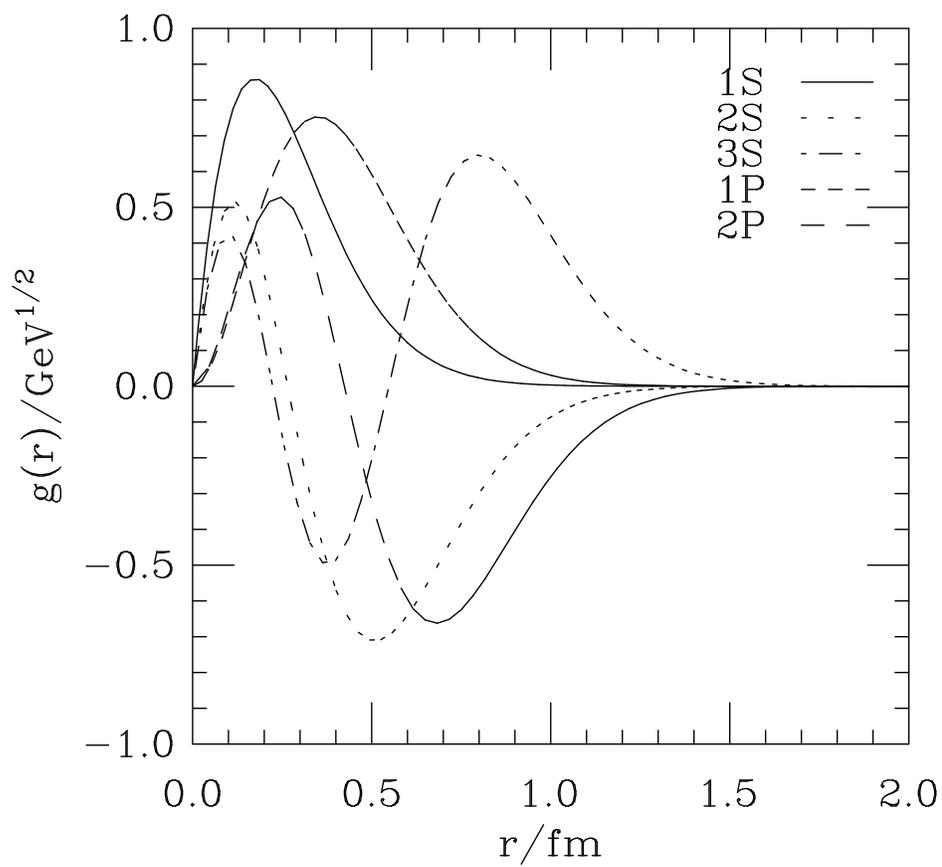}}

\end{center}
\caption{Bottomonium wave functions.}
\label{wave}
\end{figure}

\end{document}